\documentclass[aps,prl,12pt, nopacs, onecolumn, superscriptaddress, longbibliography]{revtex4-1}

\usepackage{amssymb,graphicx,color,amsmath,xfrac,bm,stmaryrd,trimclip,array,wasysym} 

\usepackage[mathletters]{ucs}
\definecolor{lightblue}{rgb}{0.17,0.39,1}
\usepackage[bookmarks, colorlinks=true, breaklinks]{hyperref}
\hypersetup{linkcolor=lightblue, citecolor=lightblue, filecolor=black, urlcolor=black}

\newcommand{\para}[1]{\left(#1\right)}
              
\newcommand{\curly}[1]{\left\{#1\right\}}

\newcommand{\fra}[2]{{#1}/{#2}}

\newcommand{\te}[1]{\text{#1}}

\renewcommand{\max}{\text{max}}
\newcommand{\model}{\text{model}} 
\newcommand{\obs}{\text{obs}}

\newcommand{\ff}[1]{{\bf #1}}
\newcommand{\B}{\text{B}}

\renewcommand{\S}{\text{S}}
\newcommand{\C}{\text{C}}
\newcommand{\N}{\text{N}}
\newcommand{\A}{\text{A}}

\renewcommand{\max}{\text{max}}

\newcommand{\Boltz}{\text{B}}
\newcommand{\Nucl}{\text{N}}
\newcommand{\Avoga}{\text{A}}

\newcommand{\CeCoIn}{$\te{CeCoIn}_5$}

\newcommand{\SuppFigCirclesandFitting}{Supplementary Fig.~1}
\newcommand{\SuppFigMoreCircels}{Supplementary Fig.~2}
\newcommand{\SuppFigACcalorimetry}{Supplementary Fig.~3}

\newcommand{\SuppFigUnshiftedToneT}{Supplementary Fig.~5}
\newcommand{\SuppFigUnshiftedCoverT}{Supplementary Fig.~6}

\newcommand{\SuppFigTalphaField}{Supplementary Fig.~8}
\newcommand{\SuppFigTalphaTwelveTesla}{Supplementary Fig.~9}

\newcommand{\SuppFigqfactors}{Supplementary Fig.~11}
\newcommand{\SuppFigPhonon}{Supplementary Fig.~12}

\newcommand{\SuppNoteNuclear}{Supplementary Note~1}

\begin{document}

\title{Magnetic field as a dynamic energy scale in quantum-critical CeCoIn$_5$}

\author{A.~Khansili}
\email{akash.khansili@gmail.com}
\affiliation{Department of Physics, Stockholm University, SE-106 91 Stockholm, Sweden}
\affiliation{Institute of Science and Technology Austria, 3400 Klosterneuburg, Austria}

\author{A.~Bangura}
\affiliation{National High Magnetic Field Laboratory, Tallahassee, Florida 32310, USA}

\author{R.~D.~McDonald}
\affiliation{Los Alamos National Laboratory, Los Alamos, New Mexico 87545, USA}

\author{K.~A.~Modic}
\affiliation{Institute of Science and Technology Austria, 3400 Klosterneuburg, Austria}

\author{B.~J.~Ramshaw}
\affiliation{Laboratory of Atomic and Solid State Physics, Cornell University, Ithaca, New York 14853, USA}
\affiliation{Canadian Institute for Advanced Research, Toronto, Ontario, Canada}

\author{A.~Rydh}
\affiliation{Department of Physics, Stockholm University, SE-106 91 Stockholm, Sweden}

\author{A.~Shekhter}
\email{Deceased}
\affiliation{Los Alamos National Laboratory, Los Alamos, New Mexico 87545, USA}

\begin{abstract}
Whether magnetic field enters the quantum-critical dynamics as a competing energy scale in strange metals remains an open question. Using thermal impedance spectroscopy, we simultaneously measure the electronic specific heat and spin-relaxation dynamics of CeCoIn$_5$. Both properties show scale invariance with temperature and magnetic field, where a single energy scale governs their crossover. Magnetic field thus competes directly with temperature to set the infrared cutoff, establishing it as a dynamic energy scale in the critical dynamics of this strange metal.
\end{abstract}

\date{\today}
\maketitle

\textit{Introduction}---Strange-metal behavior, defined by linear-in-temperature ($T$-linear) resistivity, spans a wide range of materials, including high-$T_c$ cuprates~\cite{Martin1990, Ando2004, Keimer2014}, pnictides~\cite{Kasahara2012, Analytis2014}, heavy-fermion compounds such as CeCoIn$_5$~\cite{Nakajima2004} and PuCoGa$_5$~\cite{Bauer2012}. This defining phenomenology~\cite{Anderson1988, Varma1989, Anderson1992} is often expressed in terms of a `Planckian’ relaxation rate, $\hbar/\tau = \alpha \, k_\mathrm{B}T$~\cite{Zaanen2004, Hartnoll2022}, with $\alpha$ of order unity~\cite{Homes2004, Zaanen2004, Bruin2013, Legros2019, Shekhter2026}.

The key feature underlying Planckian relaxation is the absence of an intrinsic microscopic energy scale in the quantum-critical dynamics: the critical fluctuations exist at all energy scales. In the conventional picture of quantum criticality, temperature acts as an external parameter that regulates this scale invariance: it sets an infrared cutoff, $\Lambda\sim k_\mathrm{B}T$, below which the critical fluctuations are suppressed~\cite{GellMann1954, Wilson1974}. However, temperature need not be unique in this role: any externally imposed parameter that couples to the critical degrees of freedom should likewise introduce an additional dynamic energy scale to set the infrared cutoff~\cite{GellMann1954, Wilson1974}. 

Magnetic field couples to electron spin and its orbital motion. This raises the question of whether magnetic field, too, enters the quantum-critical dynamics alongside temperature — that is, whether it plays a dynamic role~\cite{Hayes2016, GiraldoGallo2018}. This would be distinct from magnetic field acting solely as a tuning parameter of the thermodynamic phase diagram, or through the Lorentz force in Boltzmann transport near the Fermi surface~\cite{Grissonnanche2021, Ataei2022}.

To address this question, we here focus on CeCoIn$_5$~\cite{Petrovic2001}, a heavy-fermion quantum-critical metal~\cite{Bianchi2003, Ronning2005, Sakai2011, Zaum2011}. Its superconductivity is suppressed at relatively low fields, exposing the normal state over a broad range of temperatures and magnetic fields. The quantum-critical dynamics in CeCoIn$_5$ are believed to be dominated by the local $f$-electron spin fluctuations~\cite{Bianchi2003, Ronning2005, Lohneysen2007, Sakai2011, Yamashita2020}, providing a setting in which magnetic field couples directly to the critical degrees of freedom through Zeeman interaction. 

Transport measurements in high magnetic fields are strongly influenced by orbital magnetoresistance~\cite{Peierls1931, Grissonnanche2021, Ataei2022}. This question is therefore best addressed using non-transport probes. Here we use thermal impedance spectroscopy (TISP)~\cite{Khansili2023} to measure, on the same sample, the electronic specific-heat coefficient $C/T$ and the nuclear spin-lattice relaxation rate divided by temperature $1/T_1T$~\cite{Abragam1961}. Together, they provide complementary thermodynamic probes of the electronic entropy and spin dynamics associated with the same quantum-critical fluctuations. 

\begin{figure}[t!!!]
\centering
\includegraphics[width=0.7\columnwidth, keepaspectratio]{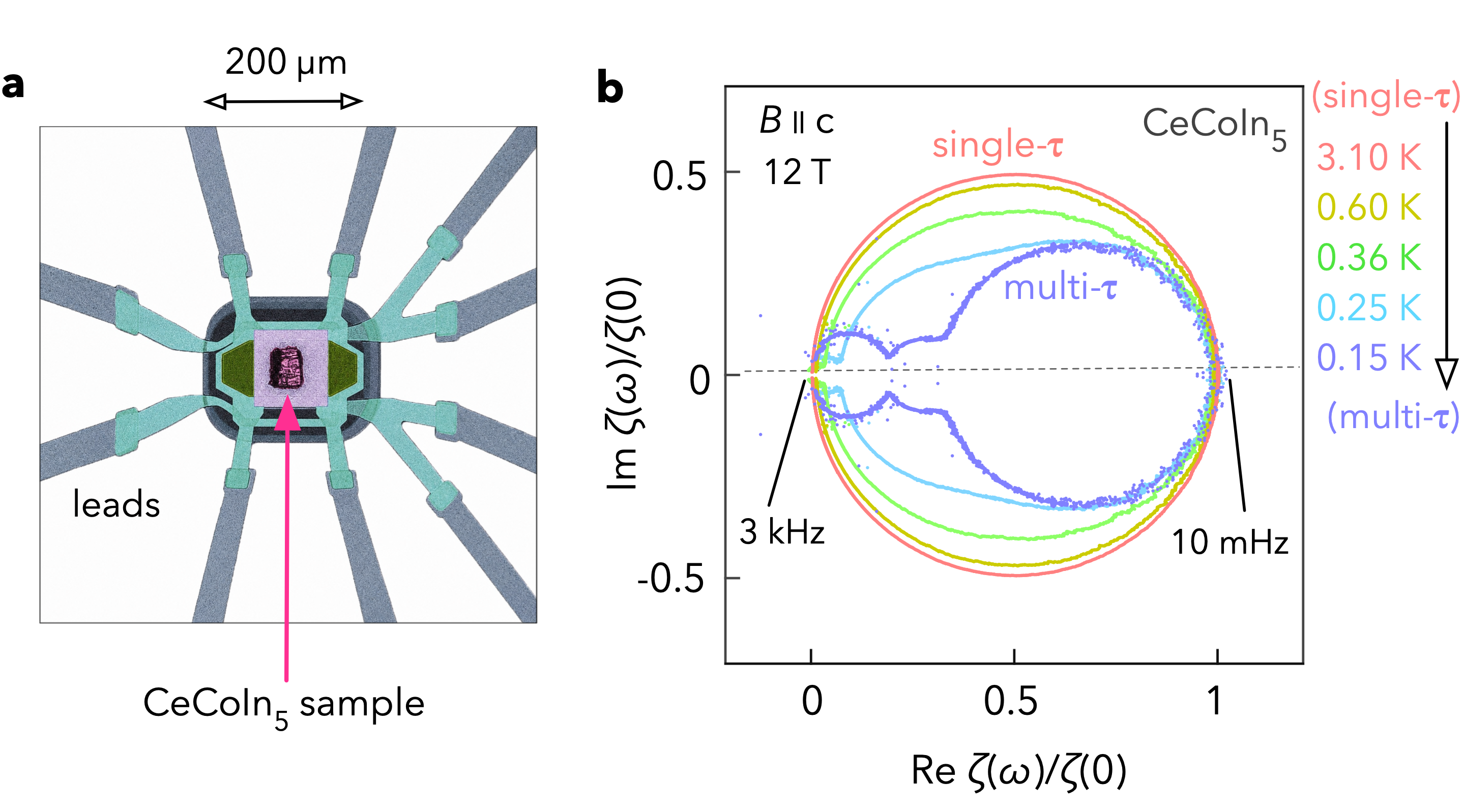}
\caption{
{Nanocalorimeter for thermal impedance spectroscopy and representative thermal impedance spectra.}
\ff{a.} Color-enhanced optical image of the calorimeter platform with the mounted CeCoIn$_5$ sample.
\ff{b.} Normalized thermal impedance spectra, $\zeta(\omega)/\zeta(0)$, of the calorimeter-sample assembly for a 12~T magnetic field applied along the $c$-axis between 0.15~K and 3.1~K. Curves below the real axis are mirrored from the measured data above the real axis for clarity. At low temperatures (e.g., the 0.15\,K curve), the spectra exhibit multiple circles, indicating multiple characteristic thermal relaxation times in the calorimeter-sample assembly (multi-$\tau$ behavior). 
}
\label{Fig:1}
\end{figure}

\begin{figure*}[t!!!]
\centering
\includegraphics[width=0.95\textwidth, keepaspectratio]{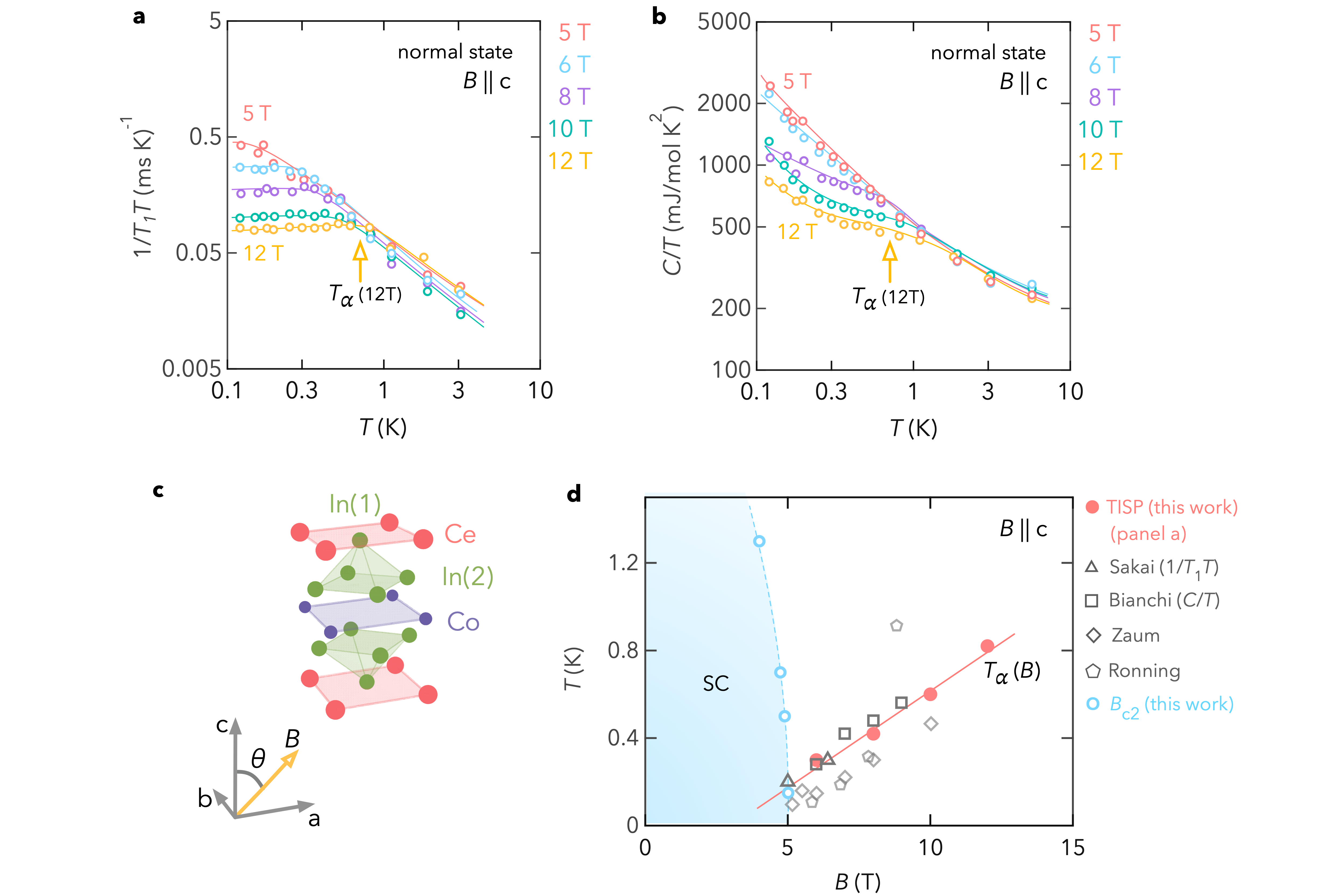}
\caption{
{Quantum-critical enhancement and magnetic  field-induced crossover for magnetic field along the $c$ axis.}
\ff{a.} Temperature dependence of $1/T_1T$ for magnetic field along the $c$ axis. The arrow marks the crossover temperature $T_{\alpha}(B\!\parallel\!c)$ at 12~T. Solid curves are guides to the eye. See {\SuppFigTalphaField}  for the determination of $T_{\alpha}(B,\theta)$ at all magnetic fields.
\ff{b.} Temperature dependence of the electronic $C/T$ for magnetic field along the $c$ axis. The arrow marks $T_{\alpha}(B\!\parallel\!c)$ at 12~T from panel~a. Solid curves are guides to the eye.
\ff{c.} Crystal structure of CeCoIn$_5$ and magnetic-field orientation convention used in this work.
\ff{d.} $T_{\alpha}(B\!\parallel\!c)$ determined by TISP (filled red circles) compared with previous measurements: NMR ($\triangle$~\cite{Sakai2011}), relaxation calorimetry ($\square$~\cite{Bianchi2003}), thermal expansion ($\diamondsuit$~\cite{Zaum2011}), and magnetoresistance ($\pentagon$~\cite{Ronning2005}). Filled blue circles denote the superconducting phase boundary determined by AC calorimetry (see {\SuppFigACcalorimetry}).
}
\label{Fig:2}
\end{figure*}



\begin{figure*}[t!]
\centering
\includegraphics[width=1\textwidth, keepaspectratio]{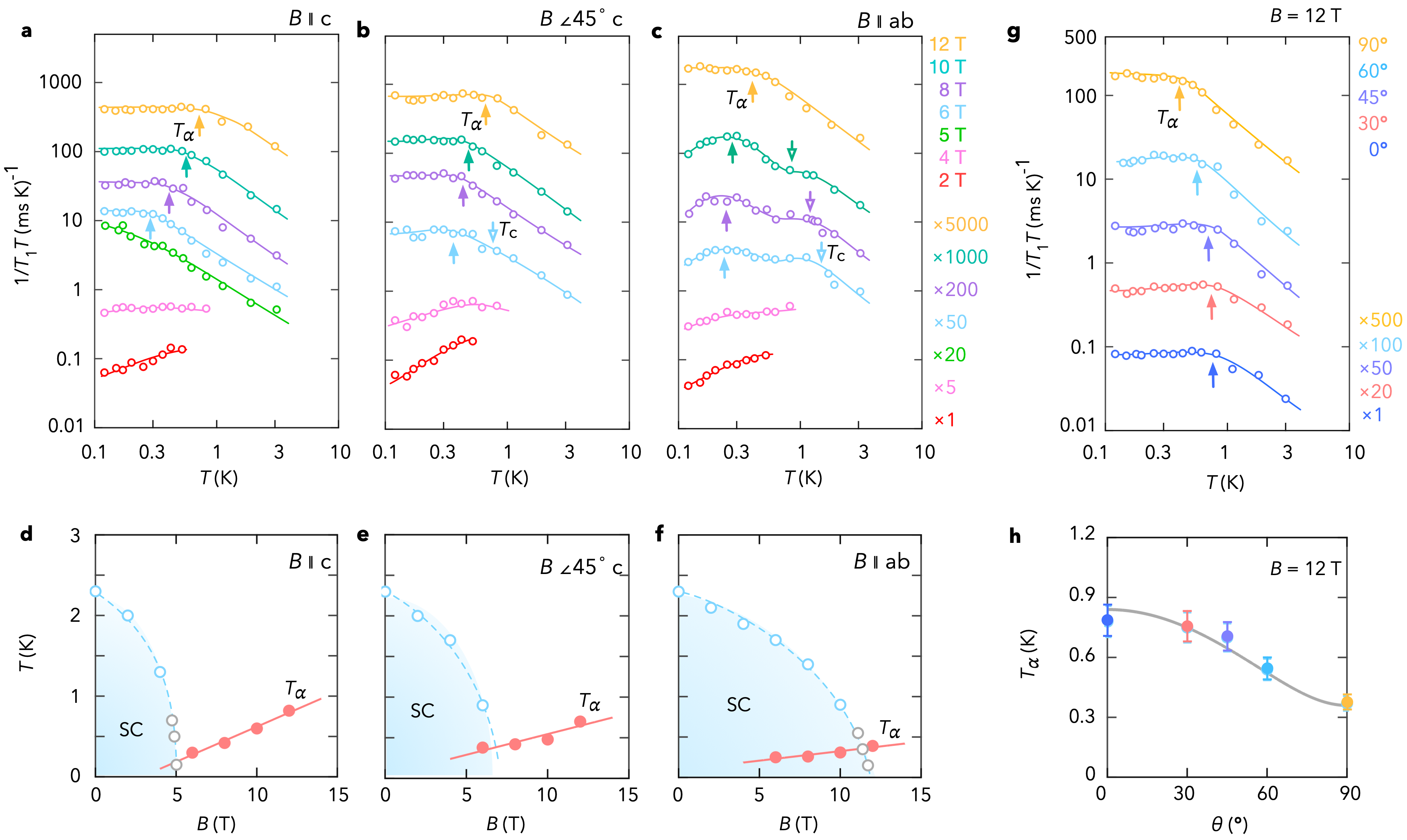}
\caption{
Crossover behavior in $1/T_1T$ for several magnetic fields and field orientations and the angular dependence of the crossover temperature $T_{\alpha}(B,\theta)$.
\ff{a,b,c.} Temperature dependence of $1/T_1T$ for fields from 2--12~T along $B\parallel c$, $45^\circ$ to $c$-axis and $B\parallel ab$ (offset vertically for clarity; see {\SuppFigUnshiftedToneT} for unshifted data). Filled arrows mark $T_{\alpha}(B,\theta)$; open arrows mark $T_{\mathrm c}$ from the electronic $C/T$ (see {\SuppFigUnshiftedCoverT}). Curves are guides to the eye (see {\SuppFigTalphaField} for the determination of $T_{\alpha}(B,\theta)$).
\ff{d,e,f.} Field dependence of $T_{\alpha}(B,\theta)$ (red markers) for the same three orientations, from panels a--c. Open blue and gray markers denote $B_{c2}(T,\theta)$ from the electronic $C/T$ ({\SuppFigUnshiftedCoverT}) and the low-temperature superconducting phase boundary from AC calorimetry ({\SuppFigACcalorimetry}), respectively. Red lines are linear fits to $T_{\alpha}(B,\theta)$.
\ff{g.} $1/T_1T$ at 12~T for several field orientations (offset vertically). Arrows mark $T_{\alpha}(12~\mathrm{T},\theta)$; curves are guides to the eye (see {\SuppFigTalphaTwelveTesla} for details).
\ff{h.} Angular dependence of $T_{\alpha}(12~\mathrm{T},\theta)$. The solid line is a fit to the lowest uniaxial angular harmonic (see {\SuppFigqfactors} for details).
}
\label{Fig:3}
\end{figure*}

\textit{Experimental details}---Thermal impedance spectroscopy determines the electronic specific heat $C$ and the effective nuclear spin-lattice relaxation time $T_1$ by analyzing the measured thermal impedance of the calorimeter-sample assembly shown in Fig.~\ref{Fig:1} over a broad frequency range~\cite{Khansili2023}. 
The nuclear spin-lattice relaxation rate obtained by TISP is an effective average over the nuclear species (see End matter) and is dominated by indium nuclei in CeCoIn$_5$. Apart from an overall scale factor, the nuclear relaxation rate obtained by TISP agrees with $^{59}$Co nuclear magnetic resonance (NMR) measurements~\cite{Sakai2011}.
Figure~\ref{Fig:2} shows $1/T_1T$ and electronic $C/T$ as determined by TISP for fields applied along the crystallographic $c$-axis, where the superconducting upper critical field is 5\,T. 
Each data point is obtained by fitting the measured thermal impedance spectra (Fig.~\ref{Fig:1}b) to Eq.~(\ref{eq:theR}) in End matter over the frequency range 10~mHz - 3~kHz. 

\textit{Results}---In the normal state of CeCoIn$_5$, at low to moderate fields, both $1/T_1T$ and the electronic $C/T$ increase strongly upon cooling, reflecting the enhancement of low-energy spin fluctuations and electronic entropy. At 5\,T, for example, the electronic $C/T$ increases by nearly a factor of 15, from about 200~mJ/mol\,K$^2$ at 6\,K to approximately 3000~mJ/mol\,K$^2$ at 0.12\,K. 
As the field increases, it produces a field-dependent crossover behavior in both $1/T_1T$ and electronic $C/T$. For fields along the $c$-axis, $1/T_1T$ crosses over from a strong temperature dependence at high temperatures to a temperature-independent value below the crossover temperature $T_\alpha(B)$ (Fig.~\ref{Fig:2}a). The electronic $C/T$ shows a change in slope at the same crossover temperature (Fig.~\ref{Fig:2}b). 

Similar crossover temperatures have been reported in $^{59}$Co NMR measurements of $1/T_1T$~\cite{Sakai2011}, relaxation calorimetry~\cite{Bianchi2003}, thermal expansion~\cite{Zaum2011}, and magnetoresistance~\cite{Ronning2005} (see Fig.~\ref{Fig:2}d). These measurements have been interpreted in terms of a magnetic field energy scale that collapses to zero near $B_{c2}$, as expected if magnetic field  tunes the system toward a quantum critical point at the superconducting upper critical field~\cite{Bianchi2003}. Instead, our analysis shows that the observed magnetic field dependence above 6~T is consistent with a finite crossover temperature $T_\alpha(B)$ at $B_{c2}$, similar to the NMR~\cite{Sakai2011} and relaxation-calorimetry measurements~\cite{Bianchi2003} (Fig.~\ref{Fig:2}d). 

The crossover temperature $T_\alpha(B,\theta)$ is observed for all field orientations. Figures~\ref{Fig:3}a-c show $1/T_1T$ for magnetic fields between 2\,T and 12\,T for field along the $c$-axis, $45^\circ$ to the $c$-axis, and for fields in the $ab$-plane. As the field is rotated toward the $ab$-plane, the superconducting upper critical field $B_{c2}$ increases toward 12\,T, allowing the crossover in $1/T_1T$ to be followed into the superconducting state. The persistence of $T_\alpha(B,\theta)$ below $T_c$ shows that the underlying critical dynamics survives the opening of the superconducting gap on the Fermi surface.

\begin{figure*}[t!!!]
\centering
\includegraphics[width=1\textwidth, keepaspectratio]{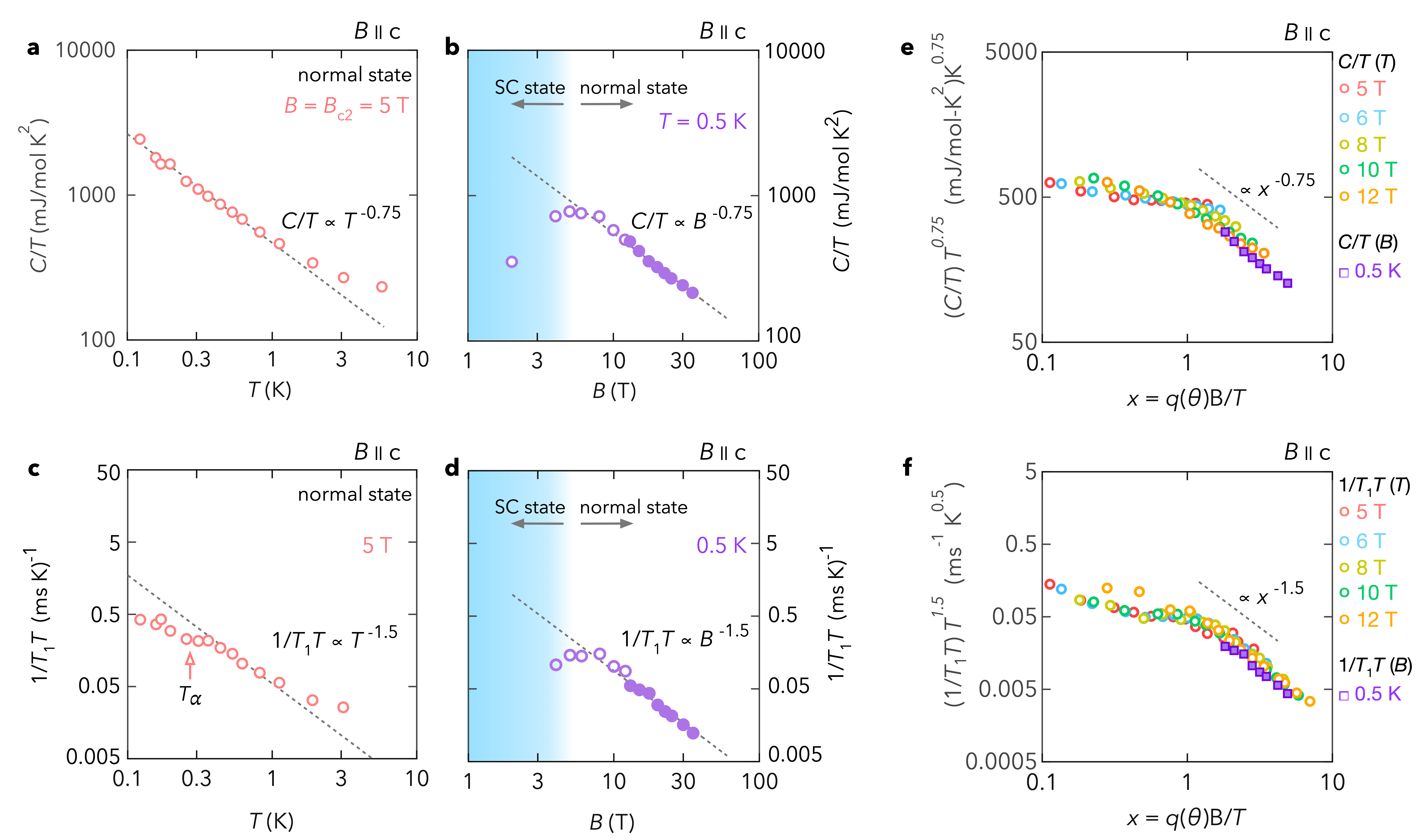}
\caption{
Temperature and field dependences of the electronic $C/T$ and $1/T_1T$ in CeCoIn$_5$.
\ff{a,b.} Temperature dependence of $C/T$ at 5~T (panel~a) and field dependence at 0.5~K (panel~b), both for fields along the $c$-axis. Dotted lines mark the power-law behavior in the high-temperature ($T\gg q(\theta)B$) and high-field ($T\ll q(\theta)B$) limits, respectively; shading in panel~b indicates the superconducting phase.
\ff{c,d.} Temperature (panel~c) and field (panel~d) dependences of $1/T_1T$, with dotted lines marking the corresponding power-law behavior; shading in panel~d indicates the superconducting phase. The upward arrow in panel~c marks the deviation from the power-law behavior, indicating the crossover temperature $T_\alpha$ for this field. 
\ff{e,f.} Quantum-critical scaling of electronic $C/T$ (panel~e) and $1/T_1T$ (panel~f) against the scaling variable $x=q(\theta)B/T$, using temperature-dependent data at several fields along $c$ (above 0.25~K for electronic $C/T$) and field-dependent data at 0.5~K.
}
\label{Fig:4}
\end{figure*}

\begin{figure}[t!]
\centering
\includegraphics[width=0.7\columnwidth, keepaspectratio]{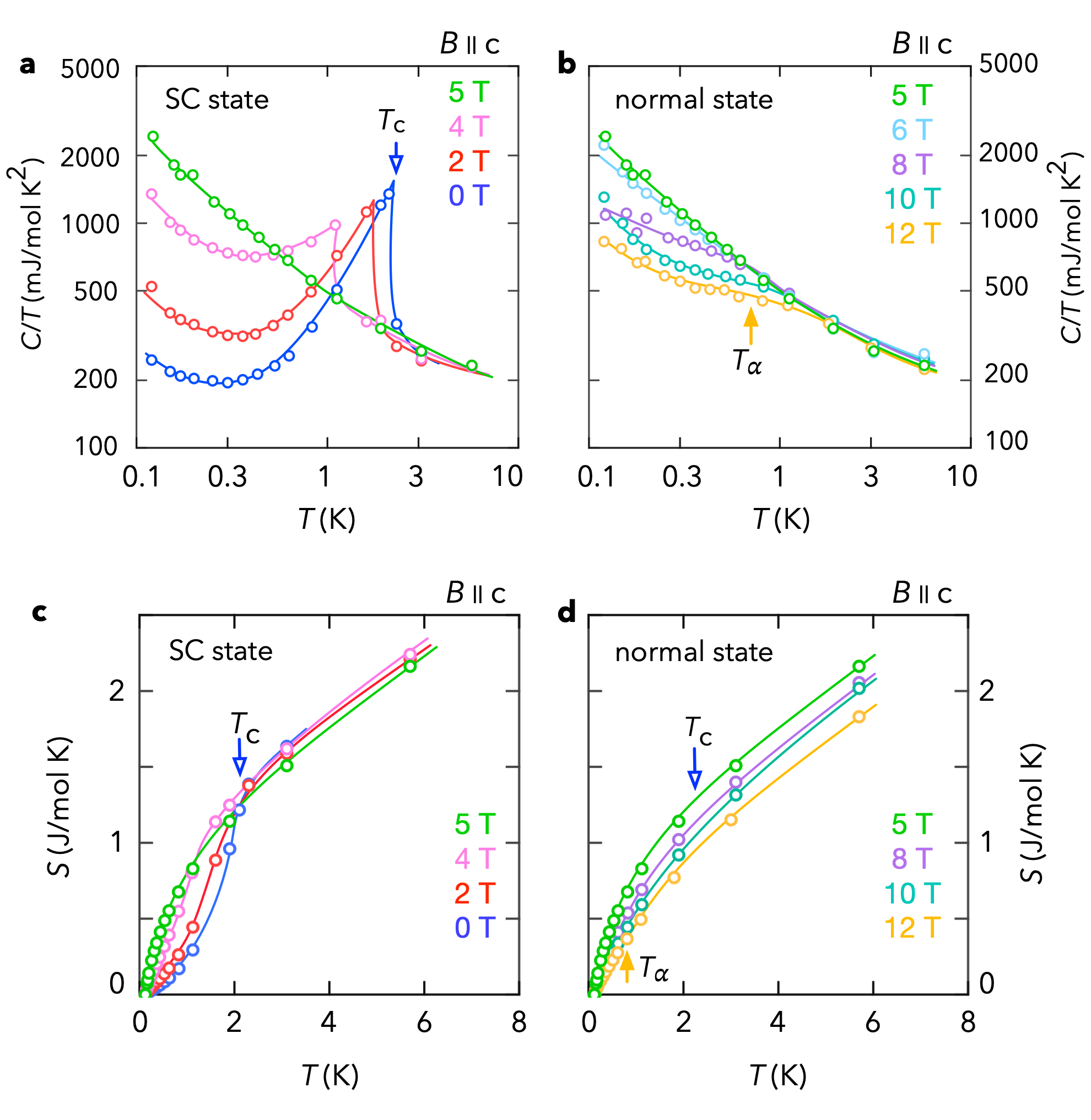}
\caption{
{Quantum critical enhancement of the electronic $C/T$ and electronic entropy in CeCoIn$_5$, and its suppression by magnetic field.}
\ff{a,b.} Temperature dependence of the electronic $C/T$ for magnetic field along the $c$-axis, below (panel~a) and above (panel~b) the superconducting upper critical field. The upward arrow marks the crossover temperature $T_{\alpha}(B,\theta)$ at 12~T determined in Fig.~\ref{Fig:3}a.
\ff{c,d.} Electronic entropy $S$, obtained by integrating the electronic $C/T$. The upward arrow marks the crossover temperature $T_{\alpha}(B,\theta)$ at 12~T determined in Fig.~\ref{Fig:3}a. 
The downward arrow marks the superconducting transition temperature at zero field, $T_c=2.3$~K. Solid curves are guides to the eye.
}
\label{Fig:5}
\end{figure}

Figures~\ref{Fig:3}d–f show the field dependence of $T_\alpha(B,\theta)$ for the three field orientations. In all cases, the crossover temperature is consistent with a linear dependence on magnetic field, $T_\alpha(B,\theta)=q(\theta)\,|B|$, with a vanishing zero-field intercept within experimental uncertainty.
Here $q(\theta)$ sets the magnitude and angular dependence of the field-induced crossover scale $T_{\alpha}(B,\theta)$. The angular dependence of $T_{\alpha}(B,\theta)$ (Fig.~\ref{Fig:3}h) is determined by combining the field-dependent measurements with the temperature dependence of $1/T_1T$ at 12\,T for several field orientations (Fig.~\ref{Fig:3}g). It is described by the lowest-order uniaxial harmonic expected for a tetragonal crystal, $q^2(\theta)=q_c^2\cos^2\theta+q_{ab}^2\sin^2\theta$, with $q_{ab}=30\pm5$~mK/T and $q_c=70\pm5$~mK/T (see {\SuppFigqfactors}). In dimensionless units, $q\,k_\mathrm{B}/\mu_\mathrm{B}$, these values are very small, 0.045 and 0.10, respectively. The anisotropy $q_c/q_{ab}\simeq2.5$ is comparable to that of $B_{c2}$, suggesting a possible connection between the quantum critical fluctuations and the superconducting anisotropy in CeCoIn$_5$.

\textit{Discussion}---The physical meaning of the $B$-linear crossover scale $T_\alpha(B,\theta)=q(\theta)B$ becomes clear when we examine the behavior of electronic $C/T$ and $1/T_1T$ above the crossover temperature (Fig.~\ref{Fig:4}). 
At high temperatures, $T\gg q(\theta)B$, the electronic specific heat follows an approximate power-law behavior $C/T\propto T^{-0.75}$ (Fig.~\ref{Fig:4}a), while the nuclear spin-lattice relaxation rate follows $1/T_1T\propto T^{-1.5}$ (Fig.~\ref{Fig:4}c). In quantum-critical metals, such power-law temperature dependencies arise because temperature sets the infrared cutoff $\Lambda$ of the critical fluctuation spectrum, $\Lambda\sim k_\mathrm{B}T$~\cite{Wilson1975}. The observed power-law behavior therefore reflects the scale invariance of electronic $C/T$ and $1/T_1T$ arising from the infrared cutoff $\Lambda$.

At high magnetic fields, $q(\theta)B\gg T$, both quantities exhibit approximately the same power-law dependencies on magnetic field: $C/T\propto B^{-0.75}$ (Fig.~\ref{Fig:4}b) and $1/T_1T\propto B^{-1.5}$ (Fig.~\ref{Fig:4}d). Because the power laws in temperature arise from temperature setting the infrared cutoff $\Lambda$, the identical power laws in magnetic field indicate that magnetic field sets the same infrared cutoff $\Lambda$ at high magnetic fields. Magnetic field therefore enters the critical dynamics as a dynamic energy scale, competing with temperature to determine the infrared cutoff, $\Lambda=k_\mathrm{B}\,\max\{T,q(\theta)B\}$. The crossover at $T_\alpha(B,\theta)=q(\theta)B$ then marks the boundary between the temperature-dominated regime, $T\gg q(\theta)B$, observed via power laws in temperature, and the field-dominated regime, $q(\theta)B\gg T$, observed via the same power laws in magnetic field. 

This dynamic competition between magnetic field and temperature is further revealed by the scaling collapse shown in Fig.~\ref{Fig:4}e,f. Plotted against the single scaling variable $x=q(\theta)B/T$, the data collapse onto universal crossover functions, $C/T\propto T^{-\alpha}g(x)$ and $1/T_1T\propto T^{-\alpha^\prime}f(x)$, with $\alpha\approx0.75$ and $\alpha^\prime\approx1.5$. Here $f(x)$ and $g(x)$ are associated quantum critical scaling functions that encapsulate the competition between temperature and magnetic field. The collapse of the $1/T_1T$ data extends over the full measured temperature range, while for electronic $C/T$, data below $0.25$~K are excluded because of the additional low-temperature entropy discussed below. Importantly, the scaling variable uses the independently determined field scale $q(\theta)$ from Fig.~\ref{Fig:3}, and critical exponents $\alpha$ and $\alpha^\prime$ from high-temperature and high-field limits in Fig.~\ref{Fig:4}a-d, rather than an adjustable parameter, demonstrating that the same field scale that governs the crossover also controls the quantum-critical scaling of both observables.

Figure~\ref{Fig:5} shows the electronic $C/T$ and the entropy obtained by integrating $C/T=dS/dT$. The upward arrow in Fig.~\ref{Fig:5}b marks $T_\alpha(B,\theta)$ determined independently from $1/T_1T$. 
At higher fields, where $T_\alpha(B,\theta)$ lies in the normal state, the electronic entropy accumulated below $T_\alpha(B,\theta)$ decreases systematically with increasing magnetic field (Fig.~\ref{Fig:5}d), consistent with progressive suppression of the low-energy critical dynamics by the magnetic-field energy scale. At low fields, the entropy obtained by integrating $C/T$, reaches the same value at $T_c$ within experimental uncertainty (Fig.~\ref{Fig:5}a,c), indicating that superconductivity redistributes the critically enhanced low-energy electronic entropy rather than removing it. This large underlying entropy then manifests itself in the unusually large specific-heat jump, $\Delta C/C\simeq4$, in CeCoIn$_5$. 

The behavior of electronic $C/T$ and $1/T_1T$ at very low temperatures and high fields reveals an important qualitative difference. While $1/T_1T$ saturates below $T_{\alpha}(B,\theta)$, the electronic $C/T$ shows only a change in slope and continues to increase upon cooling, albeit more slowly. Earlier relaxation calorimetry measurements~\cite{Bianchi2003} show similar non-saturating behavior. Such a difference shows that the low-temperature electronic entropy contains an excess entropy that is not governed by the same spin-fluctuation responsible for the enhancement of $1/T_1T$. This excess entropy and the corresponding excess electronic $(C/T)_\mathrm{excess}$ persist deep into the field-dominated regime, $T\ll q(\theta)B$, where the spin fluctuations are suppressed by the magnetic field. For example, at 12\,T, $(C/T)_\mathrm{excess}$ reaches approximately 500~mJ/mol\,K$^2$ for fields along the $c$-axis and approximately 3000~mJ/mol\,K$^2$ for fields in the $ab$-plane, indicating a pronounced anisotropy in the excess specific heat.

The excess electronic entropy indicates additional low-energy degrees of freedom beyond those associated with the local $f$-electron spin fluctuations. One possible origin is a distinct class of critical fluctuations, likely nonmagnetic and associated with the Fermi surface~\cite{Ramshaw2015a}. Because these modes couple only weakly to the magnetic field, they contribute to $C/T$ while leaving no signature in $1/T_1T$. Alternatively, the excess entropy might be associated with the Fermi surface reconstruction reported in recent de Haas-van Alphen measurements~\cite{Shishido2018}.

\begin{acknowledgments} 
\emph{Acknowledgments}---We thank M.-K.~Chan, and N.~Harrison for a critical reading of the manuscript, and A.~Kapitulnik for insightful comments. We thank A.~Thamizhavel for providing single crystals of CeCoIn$_5$. We thank K.~Schneider for image editing. 
A.K. and A.R. acknowledge support from the Swedish Research Council, D. Nr. 2021-04360. 
Resistive 35~T magnet measurements were performed at the National High Magnetic Field Laboratory, which is supported by the National Science Foundation Cooperative Agreement No. DMR-1644779 and DMR-2128556 and the State of Florida. K.A.M. and A.K. acknowledge support from the European Research Council Starting Grant 101078696-TROPIC. Work at Los Alamos National Laboratory is supported by the NSF through DMR-1644779 and DMR-2128556 and the U.S. Department of Energy. A.S. acknowledges support from the DOE/BES `Science of 100 T' grant. A.S. acknowledges the hospitality of the Aspen Center for Physics, where part of the data analysis was performed. Aspen Center for Physics is supported by National Science Foundation grant No. PHY-1607611.  
\end{acknowledgments}

\bibliographystyle{apsrev4-1}
\bibliography{refs--cecoin5}


\cleardoublepage
\section*{End Matter}
\label{Sec:Methods}

\renewcommand{\thefigure}{E\arabic{figure}}
\setcounter{figure}{0}
\renewcommand{\theequation}{E\arabic{equation}}
\setcounter{equation}{0} 

\subsection*{Sample preparation}

Single crystals of CeCoIn$_5$ were grown at Tata Institute of Fundamental Research (Mumbai) using the indium-flux method. Phase purity was verified by energy-dispersive x-ray analysis~\cite{Khansili2022}. A crystal of dimensions $50(2)\times25(2)\times20(2)$~$\mu$m$^3$ (approximately $0.20~\mu$g) was mounted on the nanocalorimeter platform~\cite{Tagliati2012,Willa2017,Khansili2023} using a thin layer of Apiezon-N grease (see Fig.~\ref{Fig:1}a).

\subsection*{Thermal impedance spectroscopy}

At low temperatures, the large nuclear heat capacity together with the nuclear spin-lattice relaxation time (e.g., $\sim20$\,ms at 1\,K) introduces an additional characteristic relaxation time into the thermal response of CeCoIn$_5$.
Thermal impedance spectroscopy (TISP)~\cite{Khansili2023} measures the frequency dependence of the thermal response over a broad frequency range to resolve the individual relaxation processes through their distinct spectral signatures.
The spectral signatures of these relaxation processes uniquely encode the electronic, nuclear, and calorimeter heat capacities, and simultaneously determine the effective nuclear spin-lattice relaxation time $T_1$, providing direct access to the low-frequency nuclear spin dynamics.

\begin{figure}[t!!!]
\centering
\includegraphics[width=0.7\columnwidth, keepaspectratio]{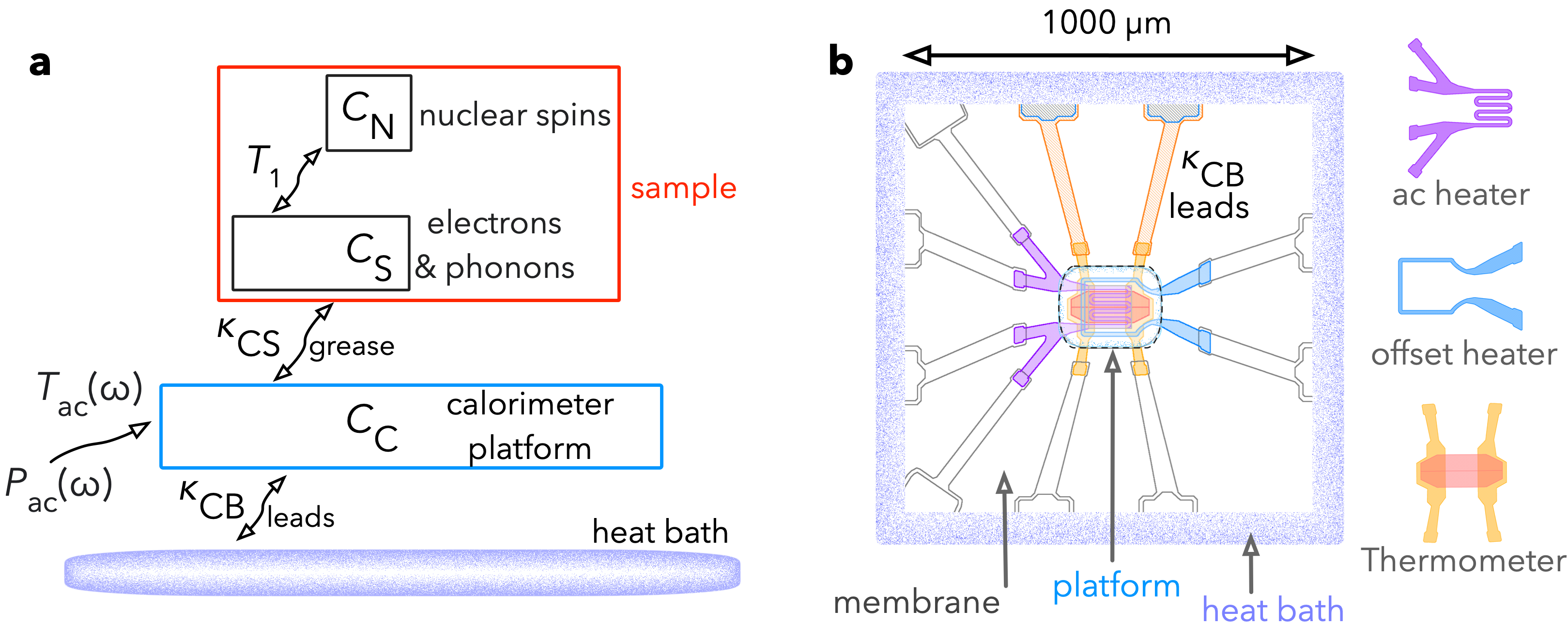}
\caption{ Components of the nanocalorimeter platform and effective heat-flow circuit of the calorimeter-sample assembly. 
\ff{a.} Thermal circuit of the calorimeter-sample assembly in this experiment.
\ff{b.} Schematic of the lithographically defined nanocalorimeter and calorimeter platform showing its components: the silicon substrate serving as the thermal bath (purple), the calorimeter platform containing the thermometer (orange) and ac heater (purple), the 150~nm SiN$_x$ membrane providing thermal isolation, and chromium leads capped with gold ( total thickness $\sim$60~nm)~\cite{Tagliati2012, Willa2017, Khansili2023}. An optical image of the calorimeter-sample assembly is shown in Fig.~\ref{Fig:1}a.
}
\label{Fig:endmatter1}
\end{figure}

\begin{figure}[b!!!]
\centering
\includegraphics[width=0.7\columnwidth, keepaspectratio]{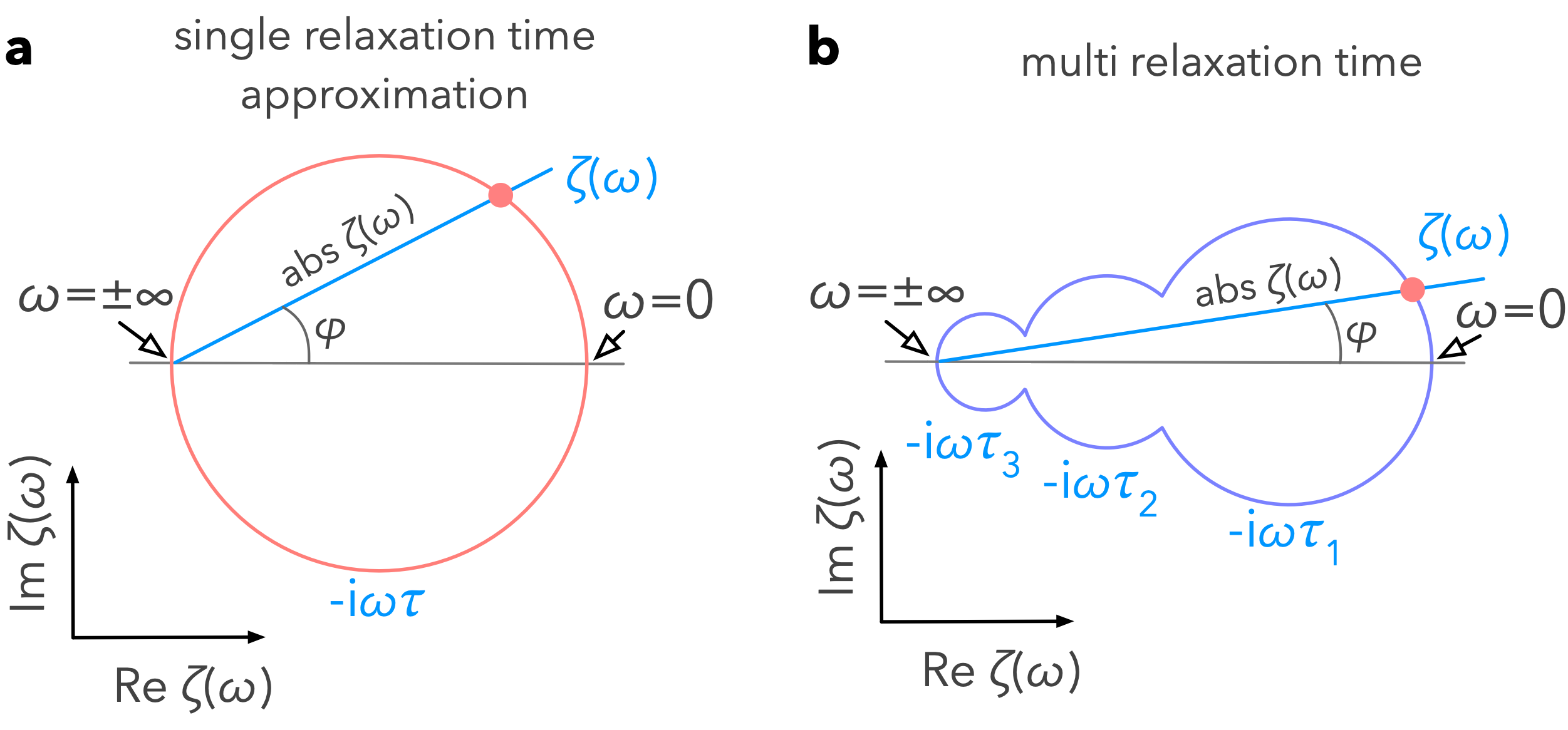}
\caption{
Complex thermal impedance $\zeta(\omega)$ in the complex plane for a single-relaxation-time response (panel~a) and a multi-relaxation-time response (panel~b). The multi-relaxation-time structure arises from multiple internal relaxation processes with characteristic relaxation times $\tau_i$. Resolving this structure requires measurements over a sufficiently broad frequency range.
}
\label{Fig:endmatter2}
\end{figure}

The thermal impedance of the calorimeter--sample assembly is
\begin{align}
\zeta(\omega)=\frac{T_{\rm ac}(\omega)}{P_{\rm ac}(\omega)}\,,
\end{align}
where $P_{\rm ac}(\omega)$ is the oscillating heating power generated by the integrated heater and $T_{\rm ac}(\omega)$ is the resulting oscillating temperature of the calorimeter platform measured by the integrated thermometer (Fig.~\ref{Fig:endmatter1}a). The heater and thermometer are lithographically integrated into the calorimeter platform~\cite{Fortune2023} (Fig.~\ref{Fig:endmatter1}b) and remain in thermal equilibrium with it over the entire measurement bandwidth. Thermal impedance spectra were measured between 10~mHz and 3~kHz using a multi-channel lock-in amplifier.
The measured thermal impedance spectra exhibit a multi-relaxation-time (multi-circle) structure in the complex plane (Fig.~\ref{Fig:endmatter2}b). Examples of thermal impedance in the complex plane are shown in Fig.~\ref{Fig:1}b and {\SuppFigMoreCircels}. 

The thermal impedance is described by the thermal circuit shown in Fig.~\ref{Fig:endmatter1}a, 
\begin{align}
\label{eq:theR}
\frac{1}{\zeta(\omega)^{\mathrm{model}}}
&= \kappa_{\mathrm{CB}} - i\omega C_{\mathrm{C}} \notag\\
&\quad
+ \frac{
-i\omega
\left(
C_{\mathrm{S}}
+
\frac{C_{\mathrm{N}}}{1-i\omega T_1}
\right)
\kappa_{\mathrm{CS}}
}{
-i\omega
\left(
C_{\mathrm{S}}
+
\frac{C_{\mathrm{N}}}{1-i\omega T_1}
\right)
+
\kappa_{\mathrm{CS}}
}.
\end{align}
where $\kappa_\mathrm{CB}$ and $\kappa_\mathrm{CS}$ are the thermal conductances between the calorimeter and bath, and between the calorimeter and sample, respectively. The heat capacities $C_\mathrm{C}$ and $C_\mathrm{N}$ correspond to the calorimeter platform and the nuclear system, respectively. The heat capacity  $C_\mathrm{S}$ is the directly determined sample heat capacity without the nuclear part.  

The observed thermal impedance spectra, $\zeta(\omega)^{\obs}$, were fit to the model $\zeta(\omega)^{\model}_{\curly{\lambda_i}}$ of Eq.~(\ref{eq:theR}) by minimizing the goodness function
\begin{equation}
\label{eq:goodness}
\begin{aligned}
g(\{\lambda_i\})_{\beta(\omega)}
&=
\int d\omega \, \beta(\omega)
\left|
\zeta(\omega)^{\mathrm{obs}}
-
\zeta(\omega)^{\mathrm{model}}(\{\lambda_i\})
\right|^2 .
\end{aligned}
\end{equation}
using a steepest descent algorithm implemented in custom software. Here $\beta(\omega)$ is a weighting function and $\{\lambda_i\}_{i=1}^6$ are the six fitting parameters of the model in Eq.~(\ref{eq:theR}). Fitting the complete thermal impedance spectrum with Eq.~(\ref{eq:theR}) simultaneously determines $C_\mathrm{S}$, $C_\mathrm{N}$, and $T_1$. Parameter uncertainties were estimated from the curvature tensor of the goodness function in parameter space, as described in Ref.~\cite{Khansili2023}. 

All results presented in the main text were obtained from unconstrained six-parameter fits. Representative fits are shown in {\SuppFigCirclesandFitting}.  The independently determined zero-field nuclear heat capacity is quantitatively consistent with the quadrupolar splitting measured by Nuclear Quadrupole Resonance (NQR) as discussed in {\SuppNoteNuclear}.

The phonon part of $C_\mathrm{S}$ is not subtracted in the main text figures. It can be estimated by fitting the zero-field total specific heat with $C_\mathrm{total}/T = \gamma + \beta T^2$ over the temperature range 7\,K to 15\,K (see {\SuppFigPhonon}). Below 3~K, the phonon contribution is less than 5\%, so the electronic $C \approx C_\mathrm{S}$, but at the highest temperature (6~K) it does affect $C_\mathrm{S}$.

\clearpage


\renewcommand{\theequation}{S\arabic{equation}}

\renewcommand{\thefigure}{S\arabic{figure}}
\setcounter{figure}{0}

\renewcommand{\theequation}{S\arabic{equation}}
\setcounter{equation}{0}


\section*{Supplementary Figures}

\begin{figure}[h!!]  
\centering 
	\includegraphics[width=0.7\textwidth, keepaspectratio]{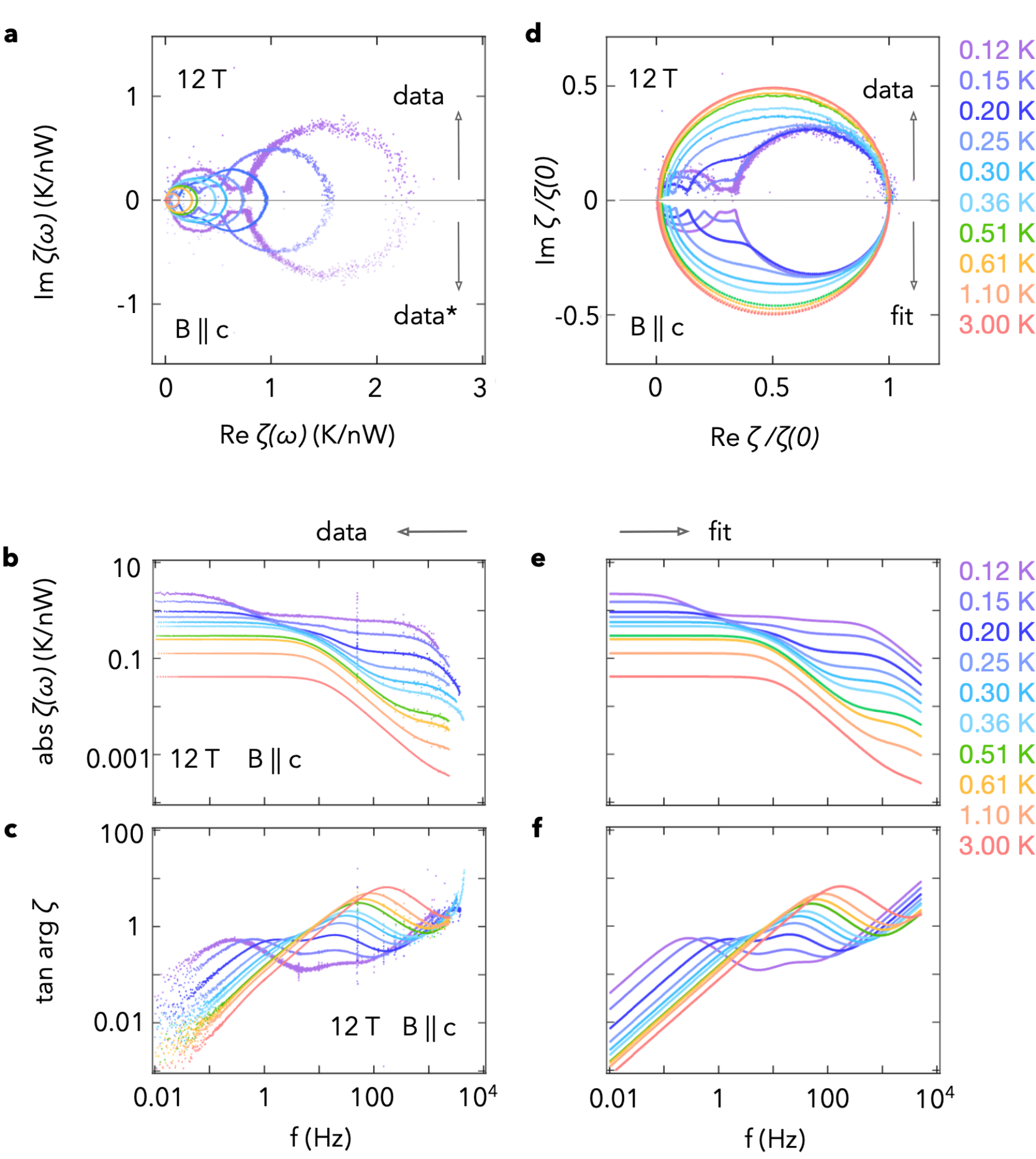} 
	\caption{\fontsize{10}{12.5}\selectfont
	{\bf Thermal impedance spectra for Fig.~1 at 12~T in the main text and their analysis using thermal impedance model in Eq.~E2. }
	{\bf a.} Thermal impedance of the calorimeter-sample assembly $\zeta(\omega)$ measured at 12~T along the $c$-axis, for temperatures between 0.12~K and 3~K. The upper half, Im$\,\zeta(\omega)>0$, shows the measured thermal impedance. The lower half, Im$\,\zeta(\omega)<0$, is the mirrored response, $\zeta(-\omega)=\zeta^*(\omega)$, shown as a guide to the eye.
	{\bf b,c.} Frequency dependence of the amplitude $|\zeta(\omega)|$ and phase $\text{arg}(\zeta(\omega))$ of the measured thermal impedance $\zeta(\omega)$ over the frequency range 10~mHz to 3~kHz.
	{\bf d.} Normalized thermal impedance, $\zeta(\omega)/\zeta(0)$. The upper half, Im$\,\zeta(\omega)>0$, shows normalized observed thermal impedance, $\zeta(\omega)/\zeta(0)$.  The lower half, Im$\,\zeta(\omega)<0$, shows fits to thermal impedance model in Eq.~E2 in the End Matter of the main text.
	{\bf e,f.} Frequency dependence of the amplitude  $|\zeta(\omega)|$ and phase $\text{arg}(\zeta(\omega))$ calculated from the thermal impedance model in Eq.~E2 using parameters obtained by fitting data in panel a. 
	}
  \label{fig:ThermImpedance} 
\end{figure}

\subsection*{Supplementary Figure~\ref{fig:ThermImpedance}} 

\clearpage

\begin{figure}[h!!!] 
\centering 
	\includegraphics[width=0.9\textwidth, keepaspectratio]{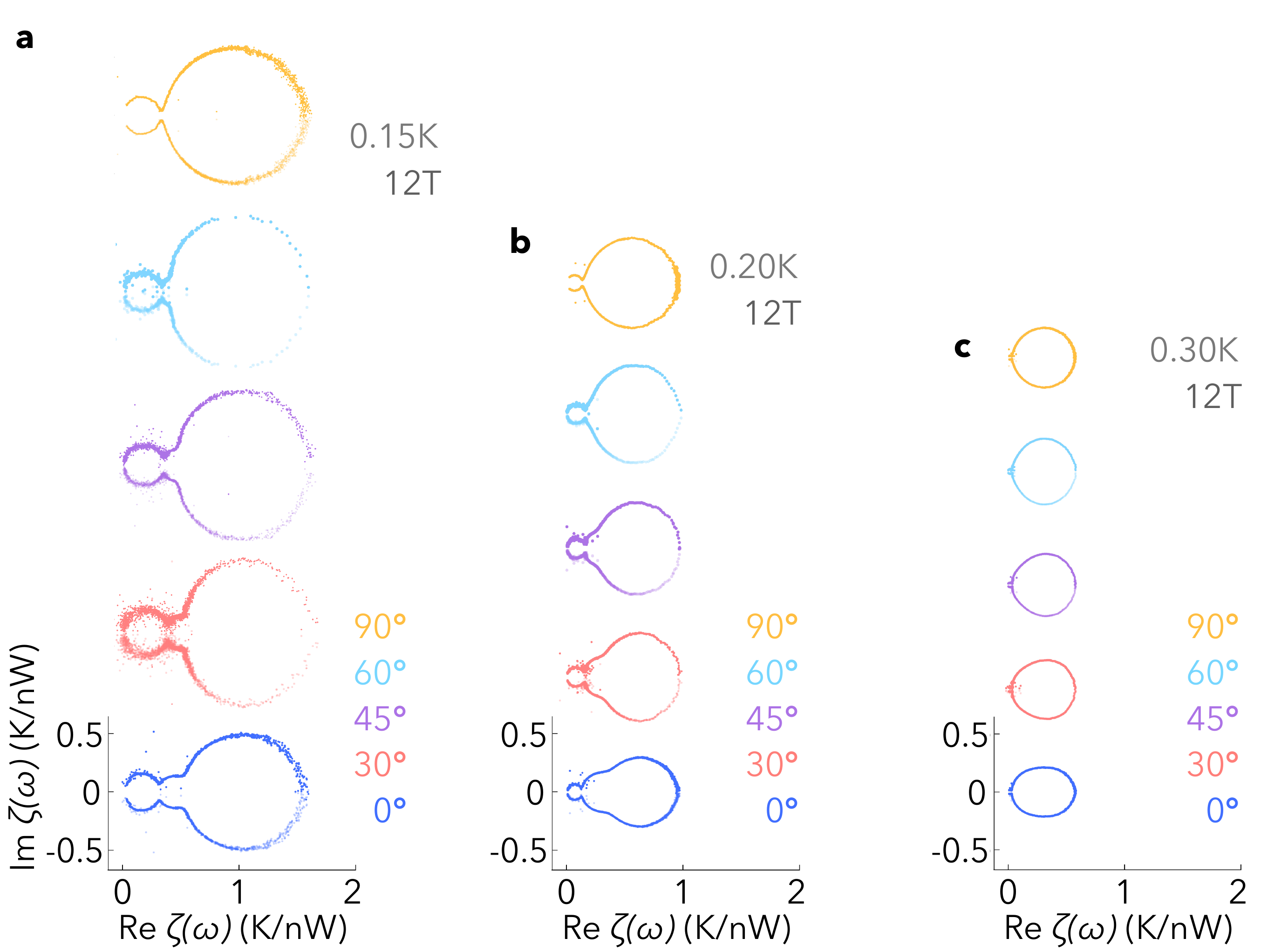} 
	\caption{
	{\bf Examples of thermal impedance spectra exhibiting multiple relaxation times.}
	{\bf a--c.} Thermal impedance spectra of the calorimeter-sample assembly measured at 12~T over the frequency range 10~mHz to 3~kHz for several field orientations at 0.15~K, 0.20~K, and 0.30~K, respectively. The data are plotted in the complex $\zeta(\omega)$ plane. The characteristic multi-circle structure reflects the multiple relaxation times that govern heat flow in the calorimeter-sample assembly over this frequency range. 
	}
\label{fig:MoreCircles} 
\end{figure}

\subsection*{Supplementary Figure~\ref{fig:MoreCircles}} 

\clearpage

\begin{figure}[h!!]
\centering   
	\includegraphics[width=0.9\textwidth, keepaspectratio]{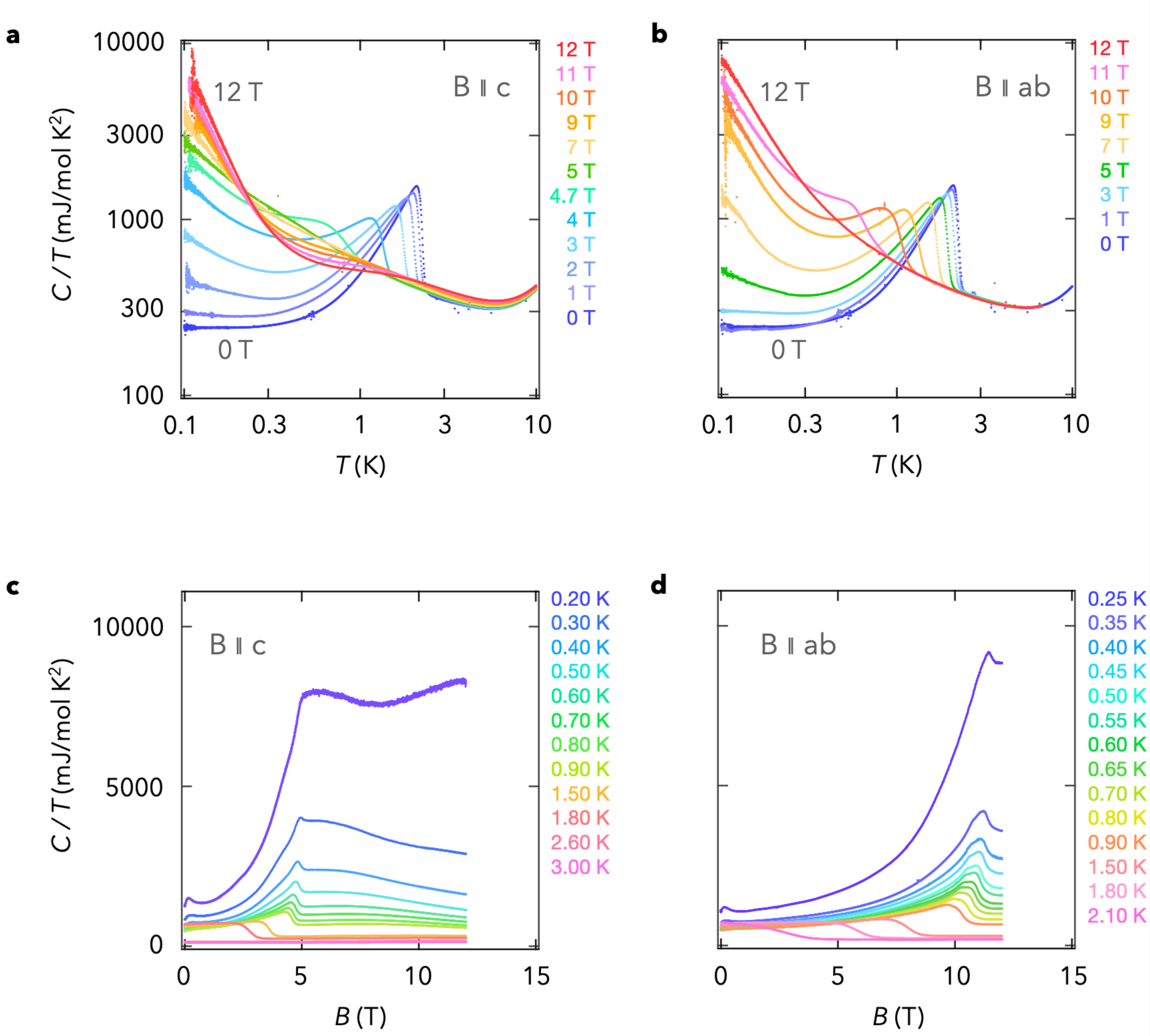} 
	\caption{ 
	\ff{ Superconducting phase boundary mapped by AC calorimetry measurements.}
	\ff{a,b.} Temperature dependence of $C/T$ for several magnetic fields applied along the $c$-axis and in the $ab$-plane.
	 \ff{c,d.} Magnetic field dependence of $C/T$ for several temperatures for the field applied along the $c$-axis and in the $ab$-plane.
	} 
  \label{Fig:ACCal} 
 \end{figure}

\subsection*{Supplementary Figure~\ref{Fig:ACCal}} 

\clearpage

\begin{figure}[h!!]
\centering
	\includegraphics[width=0.77\textwidth, keepaspectratio]{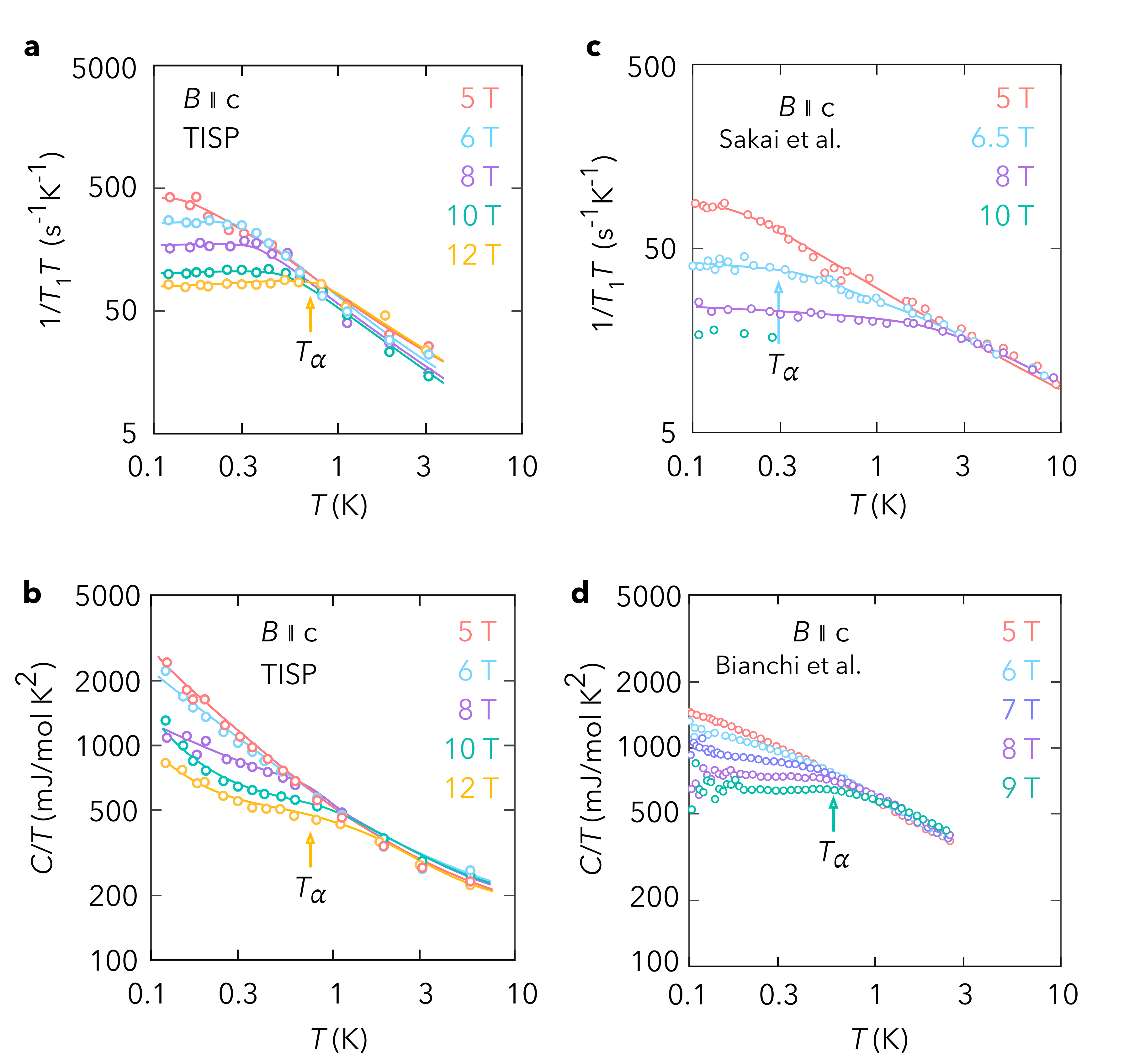}
	\caption{
	{\bf Comparison of TISP measurements of electronic $C/T$ and $1/T_1T$ with previously reported NMR and relaxation calorimetry measurements.}
	{\bf a.} Temperature dependence of $1/T_1T$ obtained by TISP for magnetic fields applied along the $c$ axis. The vertical arrow indicates the crossover temperature $T_{\alpha}(B\parallel c)$ at 12~T.
	{\bf b.} Temperature dependence of the electronic $C/T$ obtained by TISP for magnetic fields applied along the $c$ axis. The vertical arrow indicates the crossover temperature $T_{\alpha}(B\parallel c)$ determined from the corresponding $1/T_1T$ data in panel~a.
	{\bf c.} Temperature dependence of $1/T_1T$ measured by NMR~\cite{Sakai2011} for magnetic fields applied along the $c$ axis. The arrow indicates our estimate of the crossover temperature $T_{\alpha}(B\parallel c)$ at 6.5~T, used in Fig.~2d of the main text.
	{\bf d.} Temperature dependence of the electronic $C/T$ measured by relaxation calorimetry~\cite{Bianchi2003} for magnetic fields applied along the $c$ axis. The arrow indicates the crossover temperature $T_{\alpha}(B\parallel c)$ at 9~T estimated in Ref.~\cite{Bianchi2003}, used in Fig.~2d of the main text.
	}
\label{Fig:FigTwoComparison}
\end{figure}

\subsection*{Supplementary Figure~\ref{Fig:FigTwoComparison}}

\clearpage

\begin{figure}[h!!]
\centering
	\includegraphics[width=\textwidth, keepaspectratio]{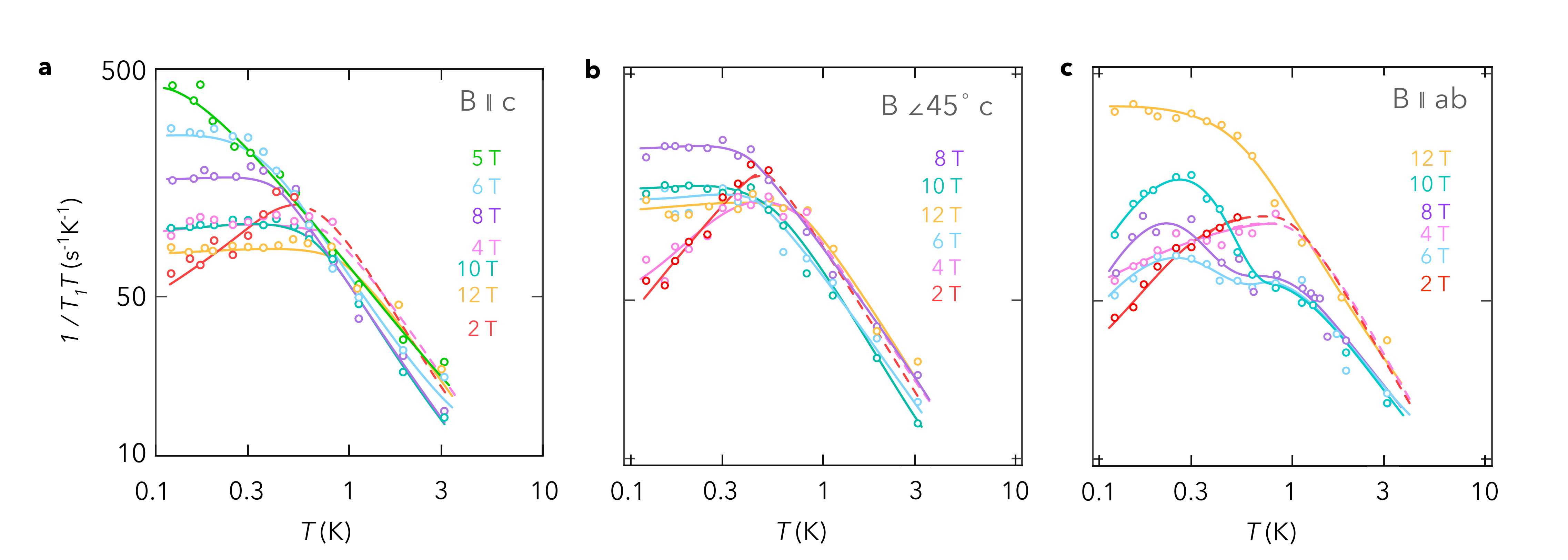}
	\caption{
	{\bf Unshifted $1/T_{1}T$ data corresponding to Figs.~3a--c of the main text.}
	{\bf a--c.} Temperature dependence of $1/T_{1}T$ at several magnetic fields for fields applied along the $c$-axis, at $45^\circ$, and along the $ab$-plane, respectively. The data are identical to those shown in Figs.~3a--c of the main text, but without the vertical offsets. Solid curves are guides to the eye.
	}
  \label{Fig:T1TUnshift}
\end{figure}

\subsection*{Supplementary Figure~\ref{Fig:T1TUnshift}}

\begin{figure}[h!!!]
\centering
	\includegraphics[width=\textwidth, keepaspectratio]{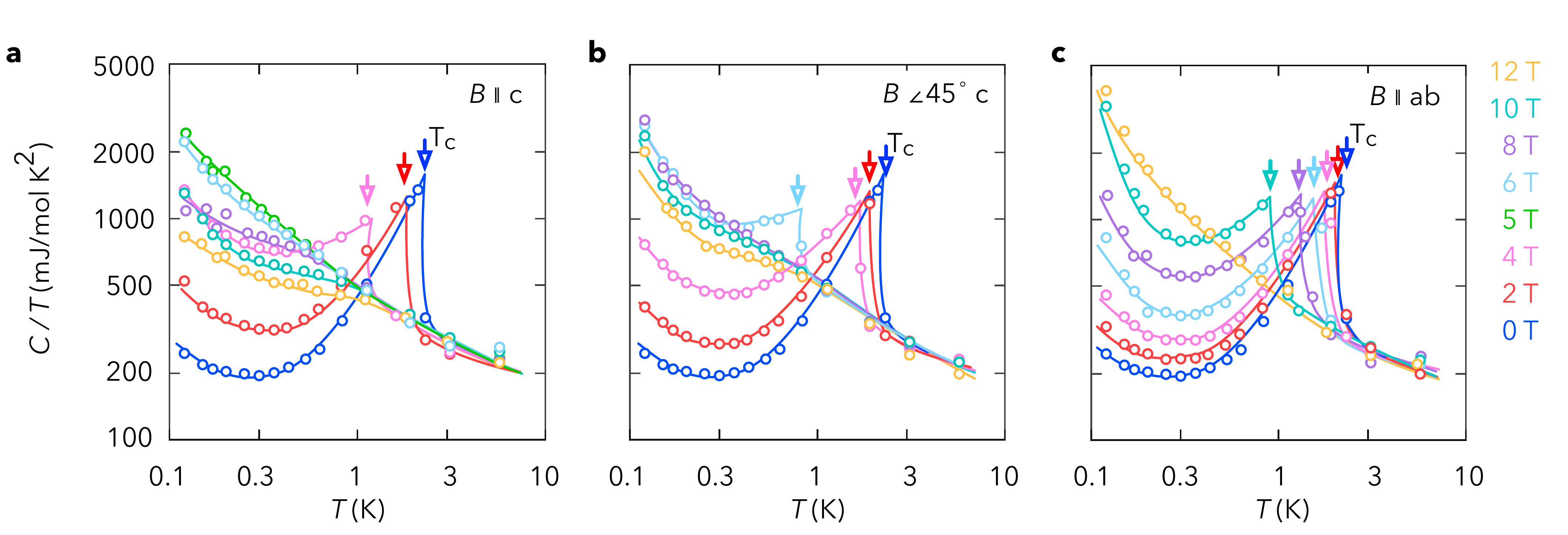}
	\caption{
	{\bf Complete data set for the electronic  $C/T$. Figures~4a--c of the main text show selected subsets of these data.}
	{\bf a--c.} Temperature dependence of the electronic $C/T$ at several magnetic fields for fields applied along the $c$-axis, at $45^\circ$, and along the $ab$-plane, respectively. The electronic $C/T$ is obtained simultaneously with the $1/T_{1}T$ data shown in Fig.~\ref{Fig:T1TUnshift} from fits to the measured thermal impedance spectra. Solid lines are guides to the eye.
	}
  \label{fig:CoverTUnshift}
\end{figure}

\clearpage
\subsection*{Supplementary Figure~\ref{fig:CoverTUnshift}}

\clearpage

\begin{figure}[h!!!]
\centering   
	\includegraphics[width=1.0\textwidth, keepaspectratio]{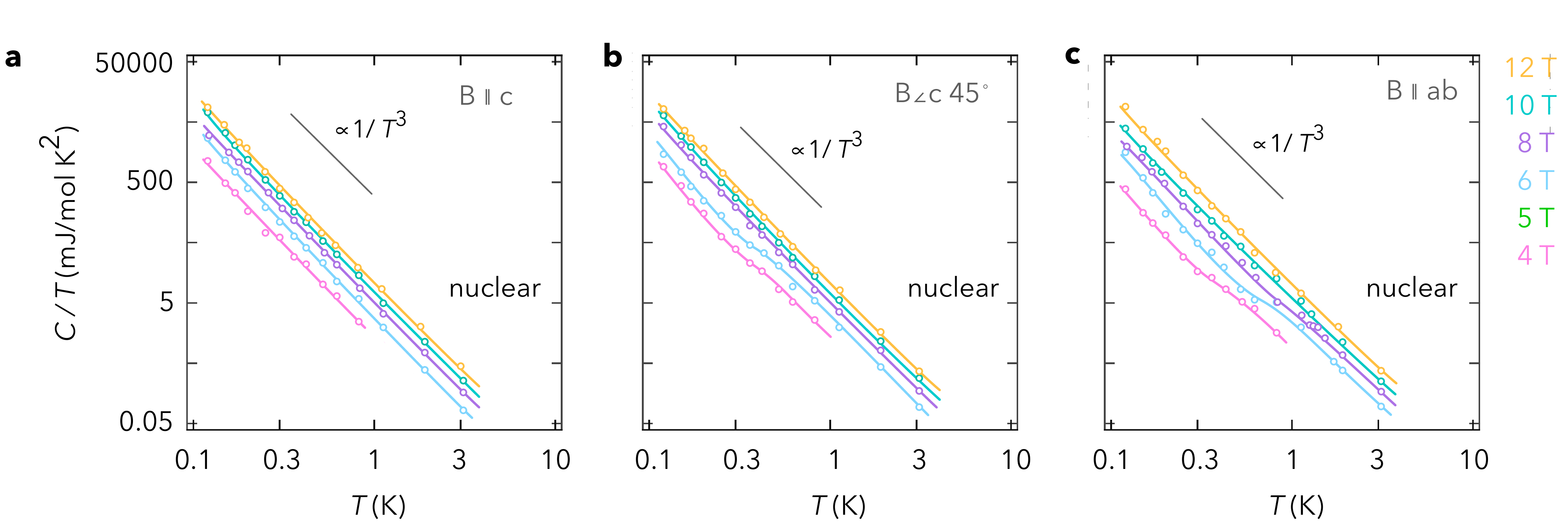} 
	\caption{ 
	\ff{Complete data set for the nuclear $C/T$ as determined by TISP simultaneously with electronic $C/T$ (Fig.~\ref{fig:CoverTUnshift}) and $1/T_1T$ (Fig.~\ref{Fig:T1TUnshift}).}
	\ff{a,b,c.} Nuclear specific heat shown as $C_N/T$ at several magnetic fields for fields applied along the $c$-axis, at $45^\circ$, and along the $ab$-plane, respectively. All colored solid curves are guides for the eye. The black solid line indicates $1/T^3$ line as a guide for eye. 
	} 
  \label{fig:NuclearSpecificHeat} 
 \end{figure}

\subsection*{Supplementary Figure~\ref{fig:NuclearSpecificHeat}} 

  \clearpage
 
\begin{figure}[h!!!]
\centering
	\includegraphics[width=0.9\textwidth, keepaspectratio]{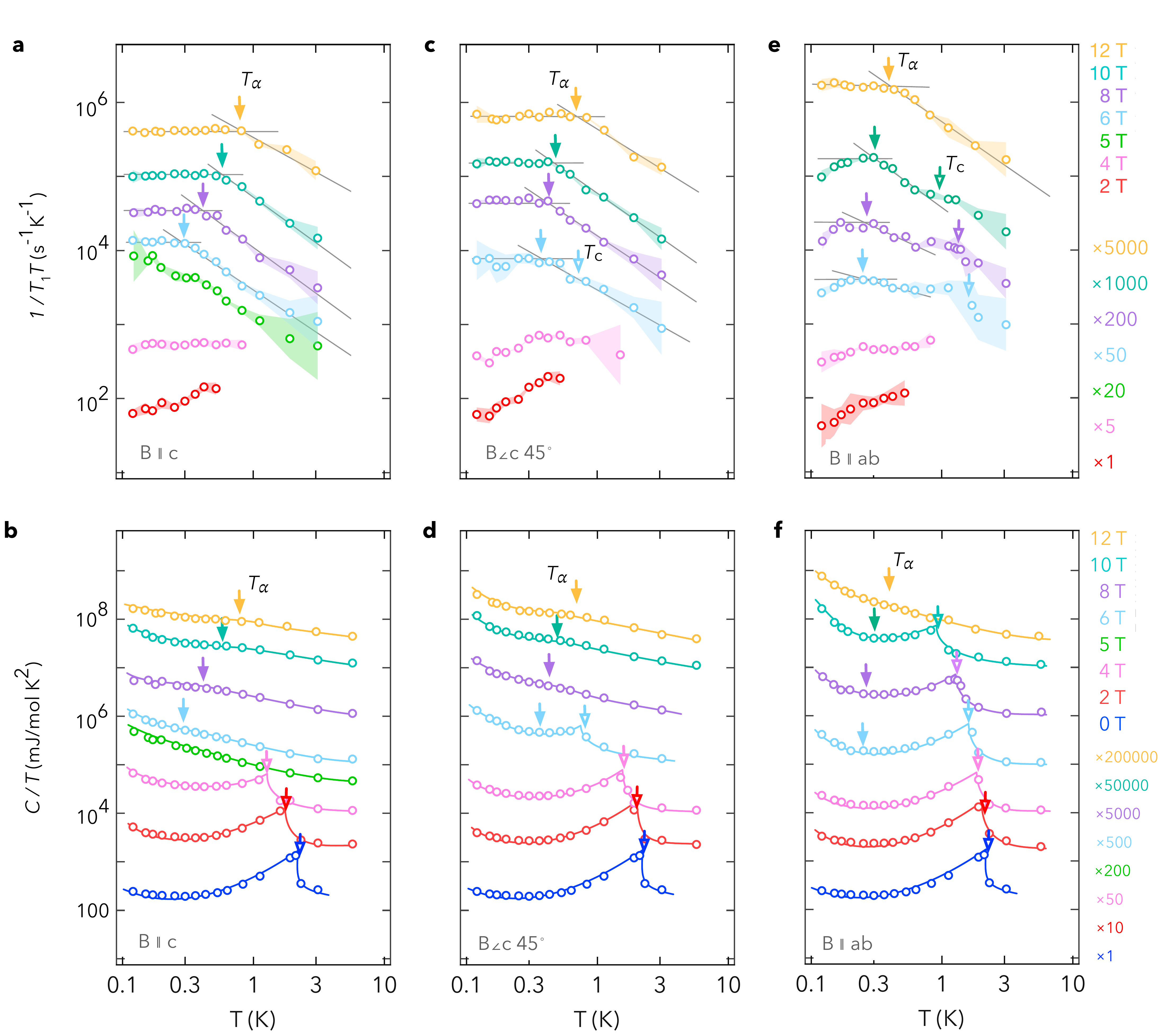}
	\caption{
{\bf Determination of $T_{\alpha}$ and $T_c$ for the complete data set of electronic $C/T$ and $1/T_1T$.}
{\bf a,c,e.} Temperature dependence of $1/T_{1}T$ at different magnetic fields for three field orientations, vertically offset for clarity (see color-coded legend). Solid gray lines indicate the limiting temperature dependence above and below the crossover. The crossover temperature $T_{\alpha}(B,\theta)$ is defined by the intersection of these limiting behaviors and is indicated by the solid downward arrow. The shaded regions indicate the uncertainty in the fitted values of $1/T_{1}T$ obtained from the thermal impedance spectra. 
{\bf b,d,f.} Temperature dependence of the electronic $C/T$ for the corresponding field orientations. Open downward arrows indicate the superconducting transition temperature $T_c$. The values of $T_{\alpha}(B,\theta)$ determined from the corresponding $1/T_{1}T$ data (panels~a,c,e) are shown as solid downward arrows.
	}
  \label{fig:TalphaField}
 \end{figure}

\subsection*{Supplementary Figure \ref{fig:TalphaField}}
 
 \clearpage

\begin{figure}[h!!]
\centering
	\includegraphics[width=0.9\textwidth, keepaspectratio]{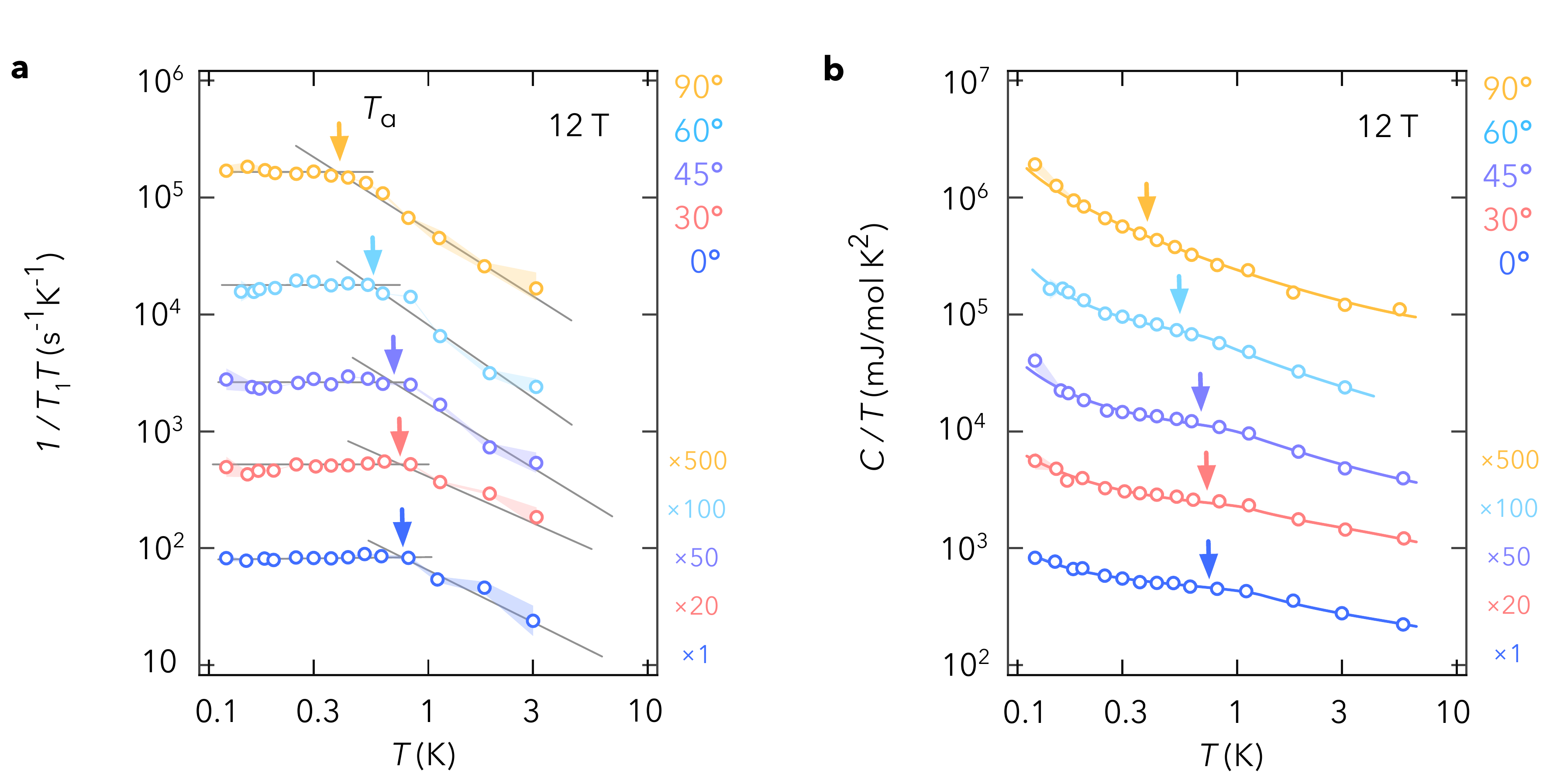}
	\caption{
{\bf Determination of $T_{\alpha}$ at 12~T for the data shown in Fig.~3g of the main text.}
{\bf a.} Temperature dependence of $1/T_{1}T$ shown in Fig.~3g of the main text, vertically shifted for clarity. Thin gray lines indicate the limiting temperature dependence below and above the crossover. The crossover temperature $T_{\alpha}$ is defined by the intersection of the two gray lines and is indicated by the vertical arrow.
{\bf b.} Temperature dependence of the electronic $C/T$ for the corresponding field orientations at 12~T, vertically shifted for clarity. The values of $T_{\alpha}$ determined from panel~a are indicated by the vertical arrows.
	}
 \label{fig:Talpha12T}
 \end{figure}

\subsection*{Supplementary Figure \ref{fig:Talpha12T}}

 \clearpage
 
\begin{figure}[h!!]
\centering
	\includegraphics[width=0.9\textwidth, keepaspectratio]{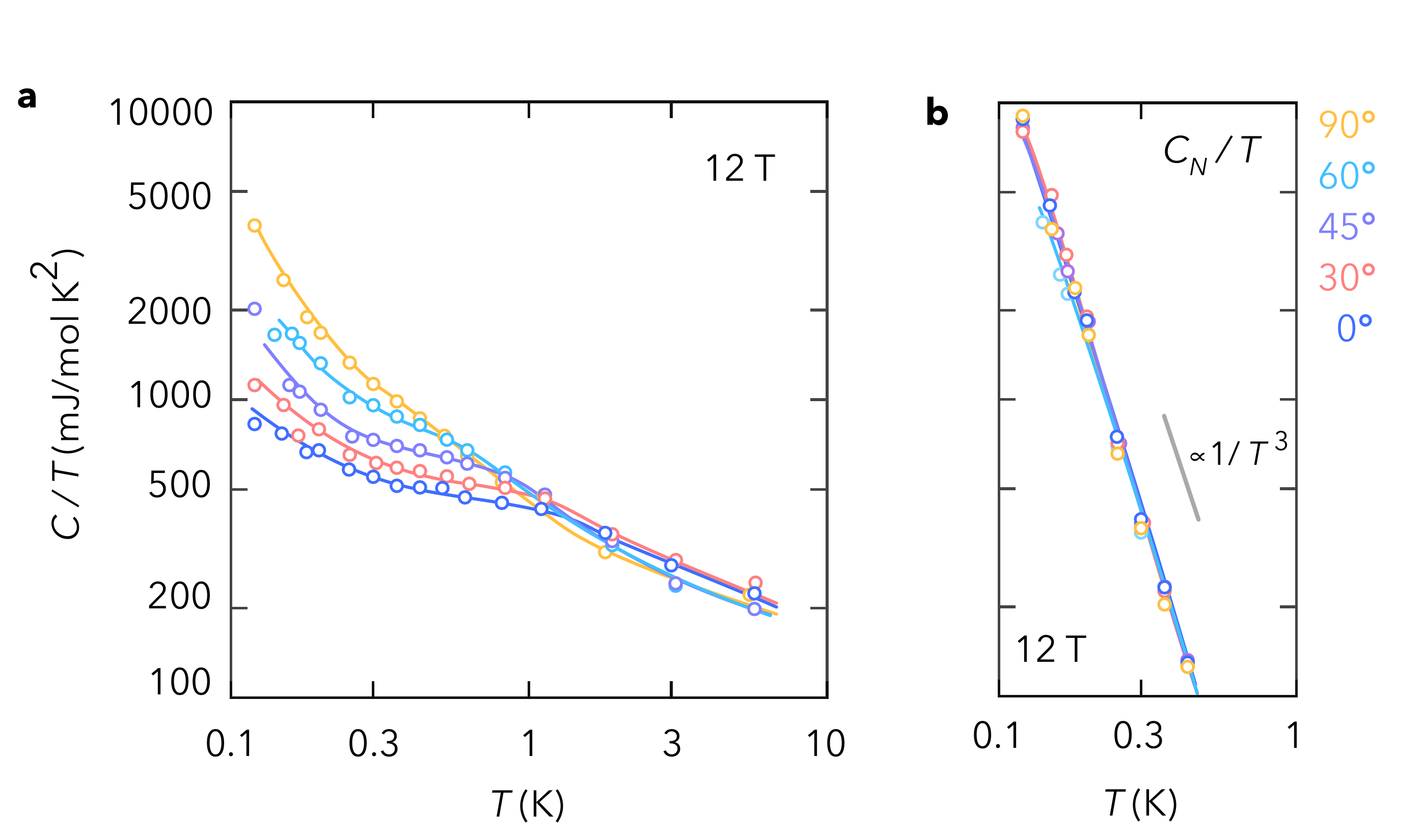}
	\caption{
	{\bf Comparison of the electronic and nuclear $C/T$ in CeCoIn$_5$ at 12~T.}
	{\bf a.} Electronic $C/T$ at 12~T for several magnetic field orientations (same data as in Fig.~\ref{fig:Talpha12T}b).
	{\bf b.} Nuclear $C_{\N}/T$, determined independently from the same TISP spectra. The black solid line indicates the  $1/T^3$ temperature dependence. Colored solid curves are guides to the eye. At 12~T, the nuclear specific heat exceeds the electronic specific heat for all field orientations. At the lowest temperature, the ratio $C_{\N}/C_{\mathrm{el}}$ reaches approximately 2 for magnetic field in the $ab$-plane and 10 for magnetic field along the $c$-axis.
		}
  \label{Fig:NuclearAngle}
\end{figure}

\subsection*{Supplementary Figure~\ref{Fig:NuclearAngle}}

\clearpage

\begin{figure}[h!!!]
\centering
	\includegraphics[width=1\textwidth, keepaspectratio]{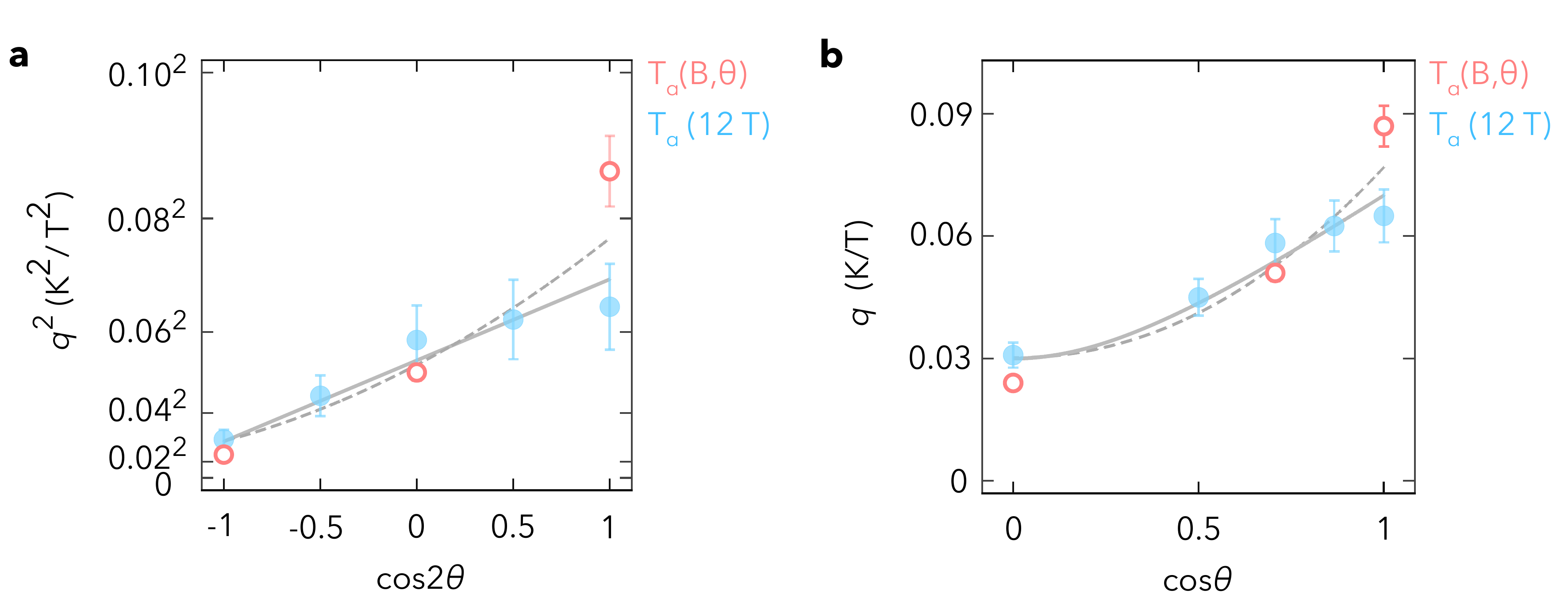}
	\caption{
	{\bf Determination of $q_c$, $q_{ab}$, and the angular dependence $q(\theta)$.}
	{\bf a.} Angular dependence of $q^2(\theta)=[T_{\alpha}(B,\theta)/B]^2$ plotted versus $\cos2\theta$. Filled blue symbols denote $T_{\alpha}(B,\theta)$ determined at 12~T from Fig.~\ref{fig:Talpha12T}. Open red symbols denote $q(\theta)$ obtained independently from the slopes in Figs.~3d--f of the main text. The solid line is a fit to the lowest angular harmonic allowed by tetragonal symmetry, $q^2(\theta)=q_c^2\cos^2\theta+q_{ab}^2\sin^2\theta$, or equivalently $q^2(\theta)=\tfrac12(q_c^2+q_{ab}^2)+\tfrac12(q_c^2-q_{ab}^2)\cos2\theta$. The fit gives $q_c=70(5)$~mK/T and $q_{ab}=30(5)$~mK/T. The dotted curve includes the next allowed harmonic, $q^2(\theta)=a+b\cos2\theta+c\cos4\theta$, resulting in only small changes in $q$-factors, $q_c=75(5)$~mK/T and $q_{ab}=25(5)$~mK/T.
	{\bf b.} The corresponding angular dependence $q(\theta)$ plotted versus $\cos\theta$. The solid curve is calculated from the lowest-harmonic fit in panel~a, while the dotted curve includes the $\cos4\theta$ correction.
	}
  \label{fig:qfactors}
\end{figure}

\subsection*{Supplementary Figure \ref{fig:qfactors}}

 \clearpage

\begin{figure}[h!!!]
\centering
	\includegraphics[width=0.6\textwidth, keepaspectratio]{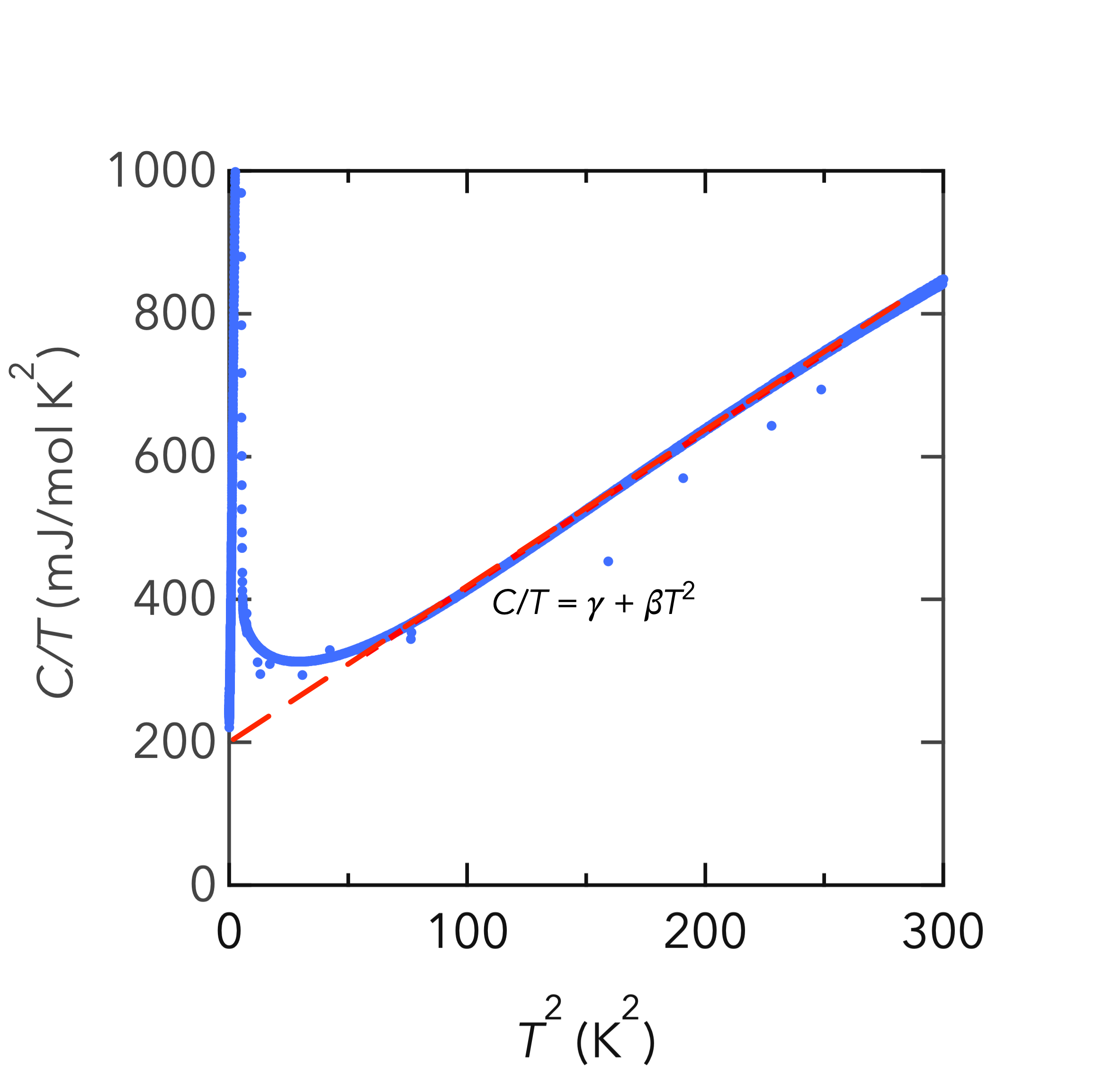}
	\caption{
	{\bf Determination of phonon specific heat at low temperatures.}
	{\bf} Zero field total specific heat obtained by ac calorimetry plotted vs $T^2$. The curve is fitted using the relation $C/T = \gamma + \beta T^2$, where $\gamma$ is the sommerfeld coefficient and $\beta$ represent the low temperature phonon contribution. The fit gives $\gamma = 200$\,mJ/mol-K$^2$ and $\beta = 2.1$\,mJ/mol-K$^4$. 
	}
  \label{fig:phonon}
\end{figure}

\subsection*{Supplementary Figure \ref{fig:phonon}}

\clearpage



\clearpage
\section*{Supplementary Notes}
\subsection*{Supplementary Note 1: \\ Electronic and nuclear specific heat due to quadrupole splitting at zero field}
\label{sec:nuclearheatcapacity}

\begin{figure}[b!!] 
\centering 
\includegraphics[width=0.9\textwidth, keepaspectratio]{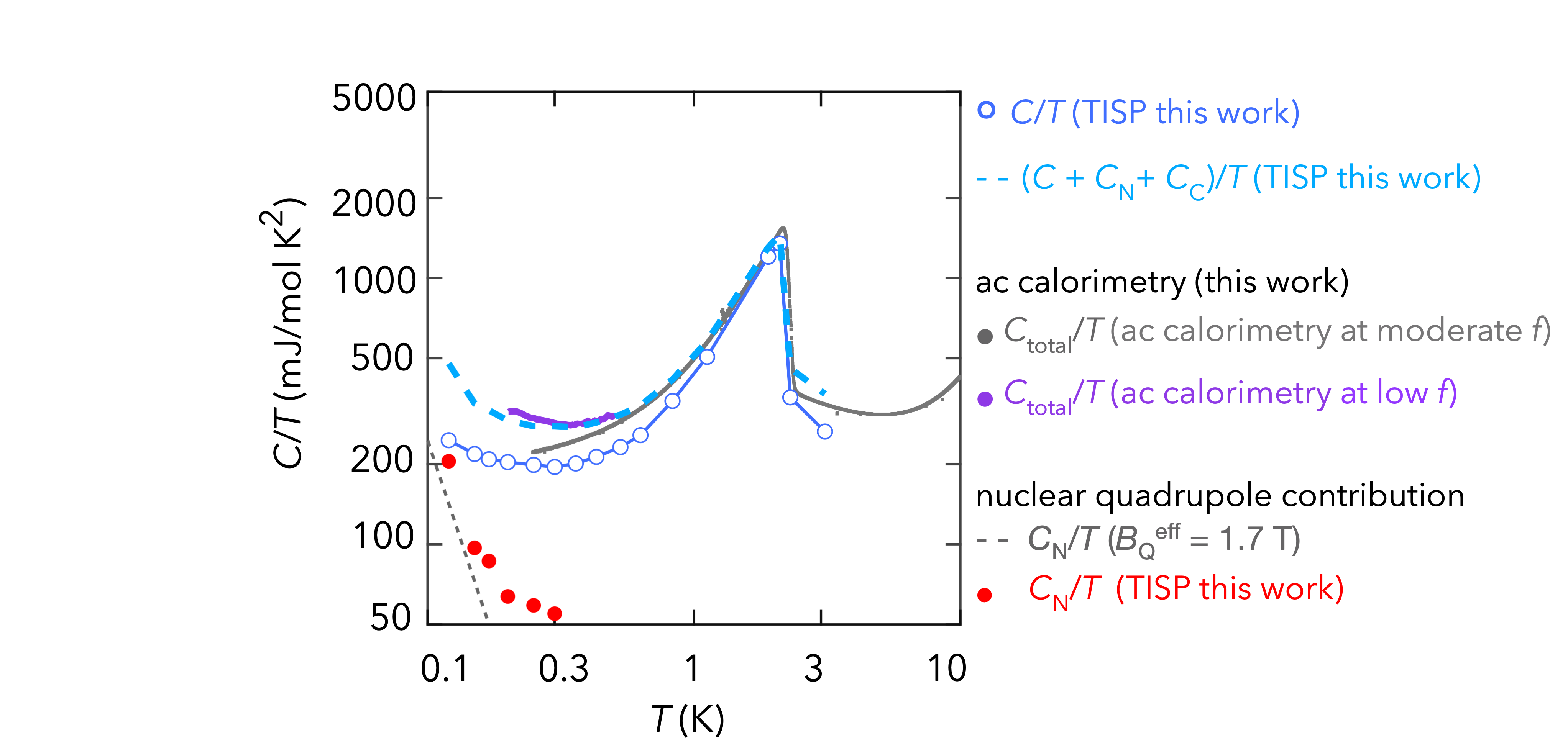} 
\caption{
{\bf Electronic and nuclear $C/T$ of CeCoIn$_5$ at zero magnetic field as determined by TISP.}
The electronic $C/T$ (open blue circles) and the nuclear $C_\N/T$ (filled red markers) determined simultaneously from TISP spectra. The gray dashed curve shows the nuclear $C_\N/T$ calculated from quadrupole splitting measured by NMR~\cite{Curro2001}. The skyblue dashed curve shows the sum of the electronic and nuclear $C/T$ (open blue circles, and filled red markers) measured by TISP (including the calorimeter platform). Black curve shows the total $C/T$ (including the calorimeter platform) measured by ac calorimetry and the purple curve at low temperature is ac calorimetry at 1~Hz frequency. As temperature decreases, the ac calorimetry frequency must be progressively decreased to capture the entire specific heat as shown in the figure.
}
\label{fig:ZeroField} 
\end{figure} 

The quadrupolar splitting in CeCoIn$_5$ measured by NMR~\cite{Curro2001} provides a model-independent prediction of the zero-field nuclear specific heat. In this Supplementary Note we calculate the corresponding high-temperature quadrupolar contribution to the nuclear specific heat (above approximately 10~mK) and compare it with the nuclear specific heat extracted from the TISP spectra. The quantitative agreement provides an independent validation of the separation of the electronic and nuclear heat capacities by TISP based solely on their distinct thermal relaxation times. Finally, the same zero-field measurements provide an opportunity to compare the TISP results with conventional AC calorimetry at very low temperatures. 

At temperatures above ten mK, the nuclear specific heat is described by the high-temperature tail of the nuclear Schottky anomaly,
\begin{align}
C_{\N} \approx & \left(\frac{B}{T}\right)^2 c_0  \,, \label{eq:nominal} \\
& c_0 =  \frac{1}{3}\,N_{\Avoga} k_{\Boltz} \left(\frac{\mu_{\Nucl}}{k_{\Boltz}}\right)^2
\sum_n  a_n \left(g_n\right)^2 I_n\,(I_n+1)
 \,, \label{eq:schottky}
\end{align} 
where $c_0$ is the reduced nuclear specific heat, i.e., its value at 1~T and 1~K. The sum in Eq.~(\ref{eq:schottky}) runs over all nuclei with nonzero spin in the unit cell, with $a_n$, $I_n$, and $g_n$ denoting their abundance, nuclear spin, and nuclear $g$-factor, respectively. Here $\mu_{\Nucl}=31.5$~neV/T is the nuclear magneton.

In CeCoIn$_5$, the reduced nuclear specific heat is determined by five $^{115}$In and $^{113}$In nuclei (which have the same nuclear spin and nearly identical nuclear $g$-factors~\cite{Stone2014}) together with one $^{59}$Co nucleus per unit cell, 
\begin{align}\label{eq:c0inCeCoIn5}
 c_0 = &  \frac{1}{3}\,N_{\Avoga} k_{\Boltz} \left(\frac{\mu_{\Nucl}}{k_{\Boltz}}\right)^2
\Bigg( 
 	5	\times (^{115\!}g)^2 \, \times \,  ^{115\!}I \,\times\, (^{115\!}I+1)  
\; + \;	1	\times  (^{59\!}g)^2 \,\times\,  ^{59\!}I  \,\times\,   (^{59\!}I+1) 
\Bigg) \notag\\
&   \approx 85\;\mu\text{JK/mol T}^2 \,.
\end{align}
where for indium $^{115\!}I=9/2$, $^{115}g=1.23$) and for cobalt $^{59\!}I=7/2$,  $^{59}g=1.32$~\cite{Stone2014}. The $^{59}$Co nucleus (second term) contributes about 13\% of the total nuclear specific heat.

Indium and cobalt nuclei also possess electric quadrupole moments, which lift the degeneracy of the nuclear spin states even in zero magnetic field. At temperatures above approximately 10~mK, the resulting quadrupolar  nuclear specific heat has the high-temperature Schottky tail, $C_{\rm Q}\propto1/T^2$. We therefore can express it in the form of Eq.(\ref{eq:nominal}) by introducing an effective quadrupolar field, $B_Q^{\rm eff}$, 
\begin{align}\label{eq:quadeff}
C_{Q} \equiv  (B_Q^{\text{eff}}/T)^2 c_0  \,\qquad 
B_{Q}^{\text{eff}}   \approx 1.7 \text{ T} \,.
\end{align}
This value represents the weighted average of the \ two inequivalent indium sites and one cobalt site in CeCoIn$_5$. At the end of this Supplementary Note we provide of derivation of the magnitude of  $B_Q^{\rm eff}$.

Figure~\ref{fig:ZeroField} compares the zero-field quadrupolar  nuclear specific heat calculated using Eq.~(\ref{eq:quadeff}) (gray dashed line) with the nuclear specific heat extracted from the TISP spectra (filled red circles). The quantitative agreement demonstrates that the separation of the electronic and nuclear $C/T$ by TISP through their temporal behavior alone is consistent with the independently measured nuclear quadrupolar splitting by NMR~\cite{Curro2001}.

At higher temperatures, where the quadrupolar nuclear $C/T$ is relatively small, the sum of the electronic and nuclear $C/T$ determined by TISP (blue dashed curve) agrees well with the AC calorimetry data (black curve). At lower temperatures, however, the nuclear spin-lattice relaxation time becomes very long in CeCoIn$_5$, and AC calorimetry must be performed at much lower frequencies to maintain thermal equilibrium between the nuclear and electronic components of the sample, which is required for accurate measurement of the total specific heat by AC calorimetry. For example, the lowest-temperature AC calorimetry data (purple curve) in Fig.\ref{fig:ZeroField} were acquired at approximately 1~Hz. At this frequency, the AC calorimetry agrees well with the total electronic and nuclear $C/T$ determined by TISP. As the temperature decreases this ac calorimetry frequency must be adjusted and decreased further to measure the entire specific heat of the calorimeter-sample assembly. We note that, unlike AC calorimetry, TISP spectra do not need to be measured at such low frequencies because the analysis explicitly accounts for the separate thermal dynamics of the electronic and nuclear subsystems.

\emph{Derivation of $B_Q^{\rm eff}$.}
Here we derive Eq.~(\ref{eq:quadeff}) starting from the nuclear quadrupolar Hamiltonian
\begin{align}
	H_Q = \frac{e^2qQ}{4I(2I-1)} (3I_z^2 - I^2)
\label{eq:quadupole}	
\end{align}
which leads to the zero-field nuclear energy levels
\begin{align}
	E_Q = \frac{e^2qQ}{4I(2I-1)} (3m^2 - I(I+1)) \,,
\end{align}
Here $Q$ is the nuclear quadrupole moment, $eq$ is the electric-field gradient at the nuclear site, $I$ is the nuclear spin, and $m=-I,\ldots,+I$ is the magnetic quantum number.

The nuclear free energy is
\begin{align}
F_Q=-k_BT\ln
\sum_{m=-I}^{I}
e^{-E_Q/k_BT},
\end{align}
from which the specific heat follows as
\begin{align}
\frac{C_Q}{T}
=
-\frac{\partial^2F_Q}{\partial T^2}\,.
\end{align}
Expanding the free energy to second order in $e^2qQ/k_BT$ for
$T\gg e^2qQ/4I(2I-1)$ gives, for a single $^{115}$In nucleus with
$I=9/2$ and a single $^{59}$Co nucleus with $I=7/2$, 
\begin{align}\label{eq:ninehalf} 
^{115}C_{Q} = \frac{11}{480}\Big(\frac{e^2\, qQ}{k_BT}\Big)^2	\,, \qquad 
	^{59}C_{Q} = \frac{3}{112}\Big(\frac{e^2\, qQ}{k_BT}\Big)^2	
\end{align}

In CeCoIn$_5$, there are two inequivalent indium sites with multiplicities 4 and 1, and one cobalt site. Therefore the nuclear specific heat (per mol) at zero field is 
\begin{align}\label{eq:quadru}
	C_{Q}^{\text{tot}} = N_{\A}k_\B \frac1{(k_\B T)^2}\Bigg( \quad 
	4\times & \frac{11}{480} \times \Big[(e^2\, qQ)_{(115,1)}\Big]^2  \notag\\
+	1\times  &\frac{11}{480} \times \Big[(e^2\, qQ)_{(115,2)}\Big]^2\notag\\
+	 1\times  & \frac{3}{112} \times \Big[(e^2\, qQ)_{(59)} \Big]^2 \quad
\Bigg)
\end{align}
where subscripts indicate the nucleus and its site. The parameter $e^2qQ$ is determined from the quadrupolar frequency,  $\nu_Q =  (\fra{6}{h})\fra{(e^2qQ)}{4I(2I-1)}$ at each nucleus and its site measured by NMR. For indium, the measured quadrupolar frequencies for two lattice sites are  $^{115}\nu_Q (1) = 8.173$~MHz,  $^{115}\nu_Q (2) = 15.489$~MHz~\cite{Curro2001}. 
This gives $e^2qQ=0.8~\mu$eV and 1.5~$\mu$eV for the two indium sites, corresponding to 9.3 mK and 17.4~mK, respectively. 
The measured quadrupolar frequency for cobalt is $^{59}\nu_Q = 0.234$~ MHz~\cite{Curro2001}. The parameter $e^2qQ$ for cobalt is 0.0135~$\mu$eV or 0.16~mK in temperature units. 
Thus, the high-temperature approximation used in Eq.~(\ref{eq:ninehalf}) is well justified over the temperature range of the present measurements (well above 10~mK).

The effective quadrupolar field $B_Q^{\rm eff}$ as the Zeeman field in Eqs.~(\ref{eq:nominal},\ref{eq:quadeff}) that produces the same high-temperature nuclear specific heat as the quadrupolar nuclear specific heat, Eq.~(\ref{eq:quadru}), 
\begin{align}
C_{Q}^{\text{tot}} \equiv  (B_Q^{\text{eff}}/T)^2 c_0  \,\qquad 
B_{Q}^{\text{eff}}   \approx 1.7 \text{ T} \,.
\end{align}

\subsection*{Supplementary Note 2: Two nuclear components in \CeCoIn}
\label{sec:twonuclearcomponents}

\begin{figure}[h!!] 
\centering 
	\includegraphics[width=0.8\textwidth, keepaspectratio]{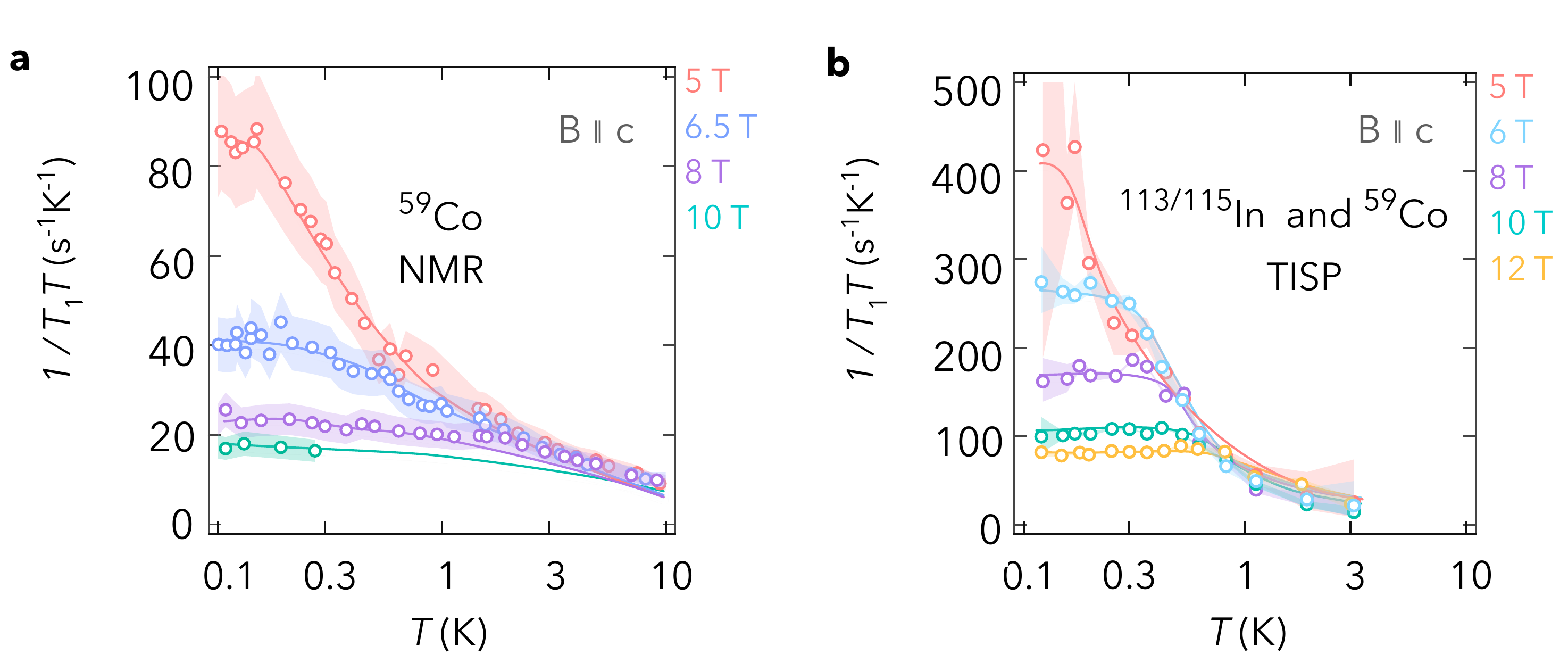} 
	\caption{ 
	{\bf NMR measurements of $1/T_{1}T$ for $^{59}$Co in {\CeCoIn} and TISP measurements of $1/T_{1}T$ for {\CeCoIn}.}   
{\bf a}.  Nuclear spin-lattice relaxation rate of $^{59}$Co in {\CeCoIn} from NMR measurements~\cite{Sakai2011} for magnetic fields along the $c$-axis. 
{\bf b}. TISP measurements of $1/T_{1}T$ for {\CeCoIn} from Fig.~2 of the main text.
	} 
	\label{fig:NMR-TISP-comparison} 
	\end{figure}

The thermal impedance model used in the main text represents all nuclear spins by a single effective nuclear subsystem characterized by a nuclear specific heat $C_{\rm N}$ and a single spin-lattice relaxation time $T_1$. In CeCoIn$_5$, however, the nuclear specific heat consists of approximately 87\% $^{115/113}$In and 13\% $^{59}$Co (Supplementary Note~1). Moreover, NMR measurements~\cite{Sakai2011} show that the spin-lattice relaxation time of $^{59}$Co is approximately five times longer than that of $^{115/113}$In (Fig.~\ref{fig:NMR-TISP-comparison}).

One might therefore naively expect the the single-nuclear-component  approximation to introduce errors of order 10\% in the extracted electronic specific heat and spin-lattice relaxation time. In this Supplementary Note we show that this is not the case. Despite the substantially different relaxation times, the error in the electronic specific heat is less than 1\%, while the error in the effective spin-lattice relaxation time is only 5--10\%. Both are smaller than the experimental uncertainties of the TISP measurements presented in this work and therefore do not affect the conclusions of the main text.

\begin{figure}[h!!] 
\centering
	\includegraphics[width=.3\textwidth, keepaspectratio]{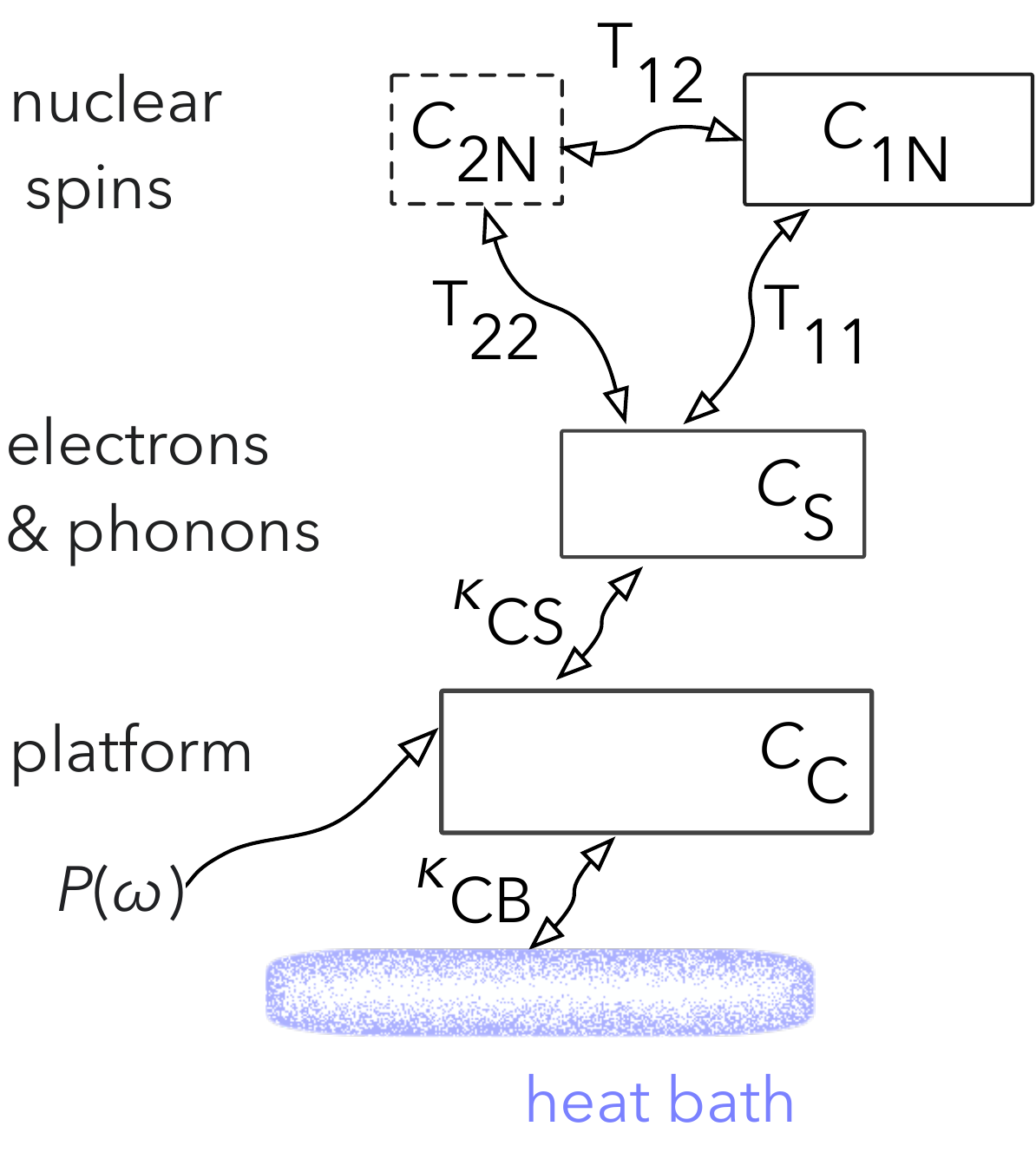} 
	\caption{{\bf Heat flow diagram of the calorimeter-sample assembly with two nuclear components.}
	{\bf a.} A sketch of the calorimeter, indicating different components.
	{\bf b.} Heat flow diagram of the calorimeter-sample assembly which determines the thermal impedance in Eq.~(\ref{eq:the2R}). {\CeCoIn} has two nuclear spin subsystems, that of $^{115/113}$In ($C_\mathrm{1N}$) and that of $^{59}$Co ($C_\mathrm{2N}$).
		} 
	\label{fig:TISP_2tau} 
\end{figure} 

The two-nuclear-species heat-flow diagram (Fig.~\ref{fig:TISP_2tau} the thermal impedance is
\begin{align}\label{eq:the2R} 
\frac1{	\zeta(\omega)_2^{\model}} =  \kappa_{\C\B}   -i \omega C_{\C} 
 + \frac{ -i \omega \! \para{ C_{\S}  + \frac{  C_{1\N} }{ -i \omega T_{11} + 1 } + \frac{  C_{2\N} }{ -i \omega T_{22} + 1 } } \kappa_{\C\S}  }{  -i \omega\!\para{ C_{\S}  +  \frac{  C_{1\N} }{ -i \omega T_{11} + 1 }  + \frac{ C_{2\N} }{ -i \omega  T_{22}+ 1 } } + \kappa_{\C\S}     } \,,
\end{align}
where $T_{11}$ and $T_{22}$ denote the spin-lattice relaxation times of In and Co respectively~\cite{Abragam1961, Hebel1963}. For simplicity we neglect cross relaxation ($T_{12}^{-1}=0$), which would only reduce the differences between the two nuclear subsystems.

To quantify the error introduced by the single-component approximation,  we fit the observed spectra using the two-nuclear-species model and using the single-nuclear-species model. For the two-nuclear-species model we use 
\begin{align}\label{eq:parameters}
	T_{11} &= T_1, \notag\\
  T_{22} &=  5 T_1 \notag\\
	C_{2\N} &=  0.13 C_{\N}\notag\\
	C_{1\N} &= 0.87 C_{\N} 
\end{align}
consistent with Supplementary Note 1 and the NMR measurements. 
We define
\begin{align}
\delta(\lambda)
=
100\,
\frac{\lambda_{\rm fit}-\lambda_{\rm true}}
{\lambda_{\rm true}},
\end{align}
where $\lambda_{\rm true}$ is the value of the corresponding parameter in the two-nuclear-species model, Eq.~(\ref{eq:the2R}), used to generate the spectrum, and $\lambda_{\rm fit}$ is obtained by fitting the same spectrum with the single-nuclear-species model, Eq.~(E2) in the End matter. 
The calculated relative errors are (see end of this Note for details) 
\begin{align}
\begin{array}{lccc}
\hline
T(\mathrm{K}) & \delta(C_S) & \delta(1/T_{1}T) & \delta(C_{\N}) \\
 & (\%) & (\%) & (\%) \\
\hline
0.12 & 0.2   & 7 & 5 \\
0.35 & 0.03  & 6 & 7 \\
1.1  & 0.001 & 5 & 7 \\
3.0  & 0.001 & 4 & 8 \\
\hline
\end{array}
\label{table:etaerrors}
\end{align}
Although the cobalt nuclei contribute 13\% of the total nuclear specific heat and relax five times more slowly than the indium nuclei, the error in the extracted electronic specific heat remains below 1\% over the entire temperature range, while the error in the effective spin-lattice relaxation time is only 5–10\%. These errors are much smaller than the changes in the nuclear subsystem itself. The remainder of this Supplementary Note explains the algebraic origin of this orthogonality.

\textit{Linear algebra of multiple nuclear species.}
It was shown above that replacing two nuclear species by a single effective one produces small changes in the fitted electronic specific heat and spin-lattice relaxation time. Here we explain why the errors in Eq.~(\ref{table:etaerrors}) remain small despite the substantial modification of the nuclear subsystem. 

Let the measured thermal impedance spectrum be $Z(\omega)$ and the model spectrum be $X(\omega)_{\lambda_i}$, where $\lambda_i$ denote the fitting parameters. Both are vectors in the linear space of functions of frequency with the scalar product
\begin{align}
\langle A(\omega)|B(\omega)\rangle
= 
\int d\omega\,
\beta(\omega)\,
A^*(\omega)B(\omega),
\end{align}
where $\beta(\omega)$ is the weighting function used in the fit. The goodness function is
\begin{align}
g(\{\lambda_i\})
=
\langle
Z(\omega)-X(\omega)_{\lambda_i}
|
Z(\omega)-X(\omega)_{\lambda_i}
\rangle .
\end{align}

Suppose that the observed spectrum cannot be described exactly by the model $X(\omega)_{\lambda_i}$. This is the case, for example, when the true spectrum is generated by the two-nuclear-species model but is fitted using the single-nuclear-species model. The difference can be written as 
\begin{align}
Z(\omega)
\rightarrow
Z(\omega)+a\eta(\omega),
\end{align}
where $a$ is small numeric factor and $\eta(\omega)$ describes the deviation from the model $X(\omega)_{\lambda_i}$. The question is how this perturbation changes the best-fit parameters $\lambda_i$ if we nevertheless fit the data using the approximate model $X(\omega)_{\lambda_i}$. 

Small changes of the fitting parameters ${\lambda_i}$ modify the model spectrum $X(\omega)_{\lambda_i}$ according to
\begin{align}
\delta X(\omega)
=
\sum_i
V_i(\omega)\,
\delta\lambda_i,
\qquad
V_i(\omega)
=
\left.
\frac{\partial X(\omega)_{\lambda_i}}
{\partial\lambda_i}
\right|_{\lambda_i=\lambda_i^0}.
\end{align}
The six functions $V_i(\omega)$ span the tangent space of spectral changes generated by small variations of the fitting parameters about the best-fit values $\lambda_i^0$. 
The function $\eta(\omega)$ can be decomposed into two orthogonal components,
\begin{align}
\eta(\omega)
=
\eta_{\parallel}(\omega)
+
\eta_{\perp}(\omega),
\end{align}
where $\eta_{\parallel}(\omega)$ lies in the tangent space spanned by the functions $V_i(\omega)$,
\begin{align}
\eta_{\parallel}(\omega)
=
\sum_i
\eta_i
V_i(\omega),
\end{align}
while $\eta_{\perp}(\omega)$ is orthogonal to that space,
\begin{align}
\langle
\eta_{\perp}
|
V_i
\rangle
=
0
\qquad
\text{for all }i.
\end{align}

Since the basis vectors $V_i(\omega)$ are generally not orthogonal, the expansion coefficients are
\begin{align}
\eta_i
=
K_{ij}
\langle
\eta
|
V_j
\rangle,
\qquad
K_{ij}
=
\left(
\langle
V_i
|
V_j
\rangle
\right)^{-1},
\end{align}
where $K_{ij}$ is the inverse Gram matrix of the basis $\{V_i(\omega)\}$.

The key point is that only the parallel component $\eta_{\parallel}(\omega)$ of $\eta(\omega)$ can be absorbed into change of the fitting parameters. 
In other words, only the part of the model difference that resembles changing one or more fitting parameters can in fact change the fitted parameters. 
 The perpendicular component $\eta_{\perp}(\omega)$ only increases the residual in the goodness function without affecting the best-fit parameters. The best-fit parameters therefore change according to 
\begin{align}
\delta \lambda_i = a  \eta_i.
\label{eq:answer}
\end{align}

A substantial physical modification of the model may therefore produce only a small change in the fitted parameters if most of the corresponding spectral change is orthogonal to the tangent space. 
This is precisely the situation encountered for the two-nuclear-species model discussed in this Note. Although introducing separate In and Co nuclear subsystems substantially modifies the nuclear dynamics, most of the resulting change in the thermal impedance spectrum is orthogonal to the directions generated by varying the fitting parameters in a single-nuclear-species approximation. Consequently, the extracted electronic specific heat changes by less than 1\%, while the effective spin-lattice relaxation time changes by only 5--10\%.

\end{document}